\documentclass[twocolumn,superscriptaddress,prl,preprintnumbers,nofootinbib]{revtex4}
\usepackage{amsmath,amssymb,bm,graphicx,color,gensymb}

\begin{document}

\title{Topography of Spin Liquids on a Triangular Lattice}

\author{Zhenyue Zhu}
\affiliation{Department of Physics and Astronomy, University of California, Irvine, California
92697, USA}
\author{P. A. Maksimov}
\affiliation{Department of Physics and Astronomy, University of California, Irvine, California
92697, USA}
\author{Steven R. White}
\affiliation{Department of Physics and Astronomy, University of California, Irvine, California
92697, USA}
\author{A. L. Chernyshev}
\affiliation{Department of Physics and Astronomy, University of California, Irvine, California
92697, USA}
\date{\today}
\begin{abstract}
Spin systems with frustrated anisotropic interactions are of significant interest due to possible exotic ground states. 
We have explored their phase diagram on a nearest-neighbor triangular lattice 
using the density-matrix renormalization group 
and mapped out the topography of the region that can harbor a spin liquid.  
We find that this spin-liquid phase is continuously connected to a previously discovered spin-liquid phase
of the isotropic $J_1\!-\!J_2$ model. The two limits show nearly identical spin correlations, 
making the case that their respective spin liquids are isomorphic to each other.
\end{abstract}
\maketitle

Some of the most visionary ideas prevail despite failing the original test case they were suggested to describe.
Such is the seminal proposal of a spin liquid  (SL) as a  ground state of the nearest-neighbor (NN) $S\!=\!\frac12$ 
triangular-lattice (TL) Heisenberg antiferromagnet \cite{spinliquid}.
Although the ground state of this model  
proved to be magnetically ordered \cite{capriotti99,white07},   
the concept of spin liquid remains highly influential in a much broader context \cite{SavaryBalentsSL16,Balents10}.

\vskip -0.025cm 
The recent surge of activity 
\cite{SciRep,Chen1,Chen2,Chen3,Chen4,Chen5,Chen6,Chen7,us,Starykh,Yuesheng17,Yuesheng17a,%
muons,multiQ,MM,MM2,Kimchi,Cava16,kappa,kappa1,Ruegg,Wang17,Balents17,Balents18}  
 brings back the NN TL model  as a  
potential holy grail of spin liquids that may provide a redemption to 
the original proposal of Ref.~\cite{spinliquid}. 
In its modern reincarnation, the key players are the highly anisotropic spin interactions, 
borne out of the strong spin-orbit coupling \cite{SciRep,Chen3,Balents17}. 
This ongoing effort is also inspired by the Kitaev SL construct for the bond-dependent 
spin interactions on the honeycomb lattice \cite{Kitaev},
although without the benefit of an exact solution in the  TL geometry. 

\vskip -0.025cm 
The highly anisotropic interactions naturally emerge from a projection 
of the large magnetic moments' Hilbert space onto the manifold of low-energy pseudospin-$\frac12$ 
degrees of freedom of the rare-earth compounds \cite{GingrasRMP,Chen3}. 
Among the recently discovered TL rare-earth-based magnets, 
YbMgGaO$_4$  (YMGO)  \cite{SciRep} has received most attention.  While the debate on the 
intrinsic vs disorder-induced nature of its spin-liquid-like response is ongoing
\cite{SciRep,Chen1,Yuesheng17,us,Kimchi,kappa1},  
a  broader family of the rare-earth TL materials has also become available \cite{Cava16,Chen3}. 

\vskip -0.015cm 
Thus, it is important to provide a much needed framework to this area
by establishing the phase diagram of the most general NN TL model
with an unbiased numerical approach that goes beyond the mean-field methods that  
favor SL  by design \cite{Chen1,Chen2}.  
That should also 
settle whether extrinsic mechanisms are at work to mimic an SL behavior in the 
cases such as YMGO \cite{us}.

\vskip -0.035cm 
In this Letter, we explore the three-dimensional (3D) phase diagram of the most general NN 
model of these materials 
by using the density-matrix renormalization group (DMRG)  aided by quasiclassical analysis.
In agreement with prior numerical work \cite{us,Balents17,Wang17},
we find that the phase diagram is dominated by well-ordered states and shows no indication that anisotropic terms 
by themselves can lead to a massive degeneracy that can favor SL states.
On the contrary, most of the phase boundaries are surprisingly close to that of the classical $S\!=\!\infty$ limit, 
implying reduced quantum fluctuations and strongly gapped states due to anisotropies.

Nonetheless, we have found a likely candidate for an SL state and created its topographic map, although a
weak or a more complicated ordering \cite{multiQ} cannot be fully ruled out for much of that region.
The maximal extent of the SL phase is achieved at the isotropic limit of the bond-independent part of the model, 
questioning that anisotropies are a prime source of an SL in these systems. 

While  the local character of the $f$-shell magnetism of the rare-earth ions dictates the dominance of the 
NN interactions, experiments suggest a sizable next-NN 
coupling $J_2$ \cite{MM,MM2}. 
We find that a four-dimensional extension of the phase diagram with $J_2$ 
allows for a natural continuity of the SL state from the anisotropic TL to the  
isotropic $J_1$--$J_2$ limit 
\cite{ZhuWhite,Yigbal,Gazza,Kaneko,schwinger17,Mishmash13,Bishop,Sheng1,Sheng2,McCulloch}.
The spin-spin correlations show no transition vs $J_2$ and are nearly identical  
between these two limits, suggesting isomorphism of the corresponding SL states. 
Our study indicates that these   
SLs are either  $Z_2$ or Dirac-like \cite{Balents17,ZhuWhite,Kaneko,Yigbal,Sheng1,Sheng2,McCulloch,Mishmash13,schwinger17},
not the ``spinon metal'' SL state, argued to exist in YMGO \cite{Chen1,Chen7}. 

\begin{figure}[t]
\includegraphics[width=0.98\linewidth]{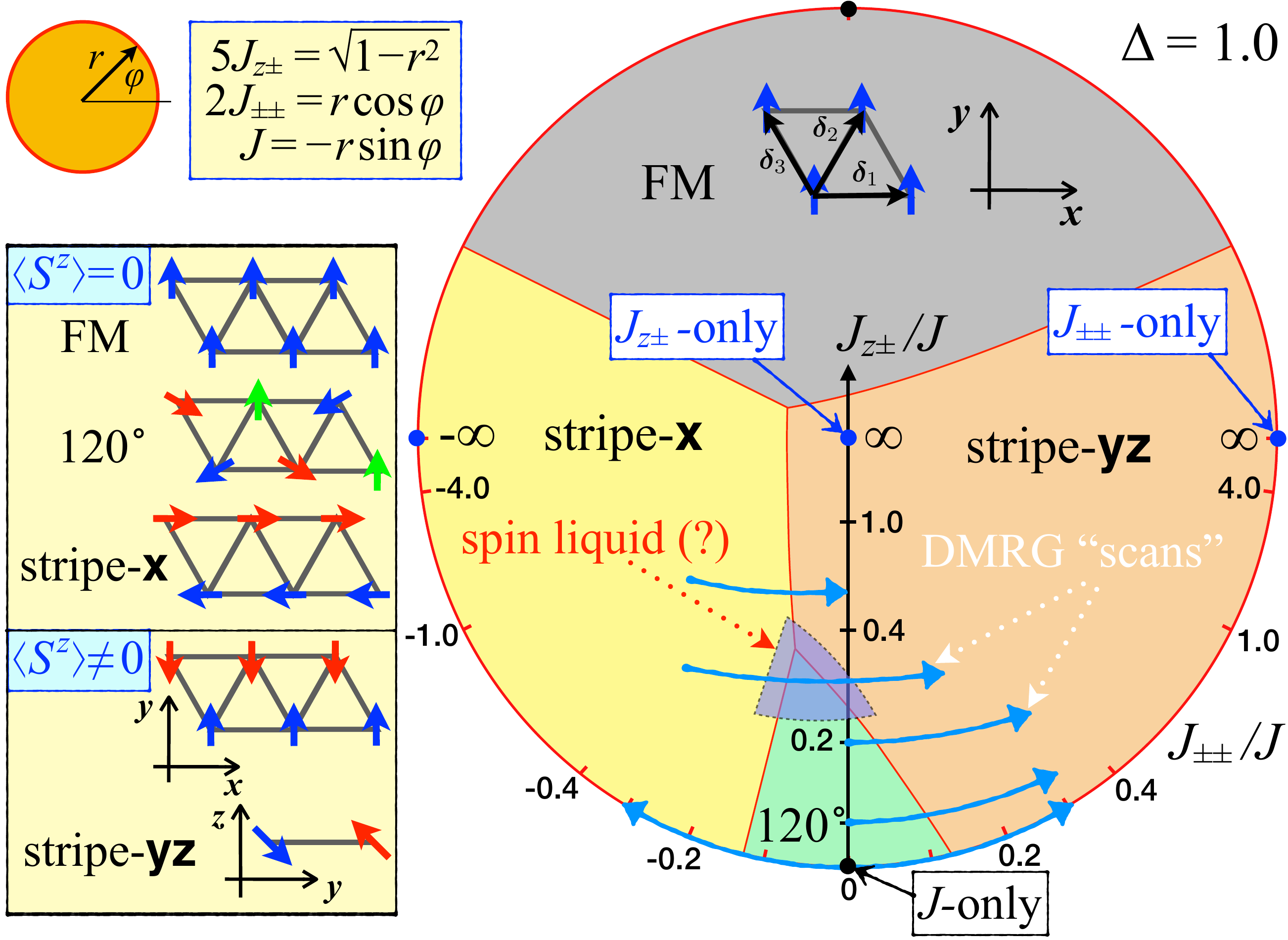}
\vskip -0.2cm
\caption{The classical 2D phase diagram of the model (\ref{H}) for  $\Delta\!=\!1.0$
using polar parametrization (see inset). The sketches of the ordered states (see text) and 
lattice primitive vectors are shown. Some of the
DMRG scans of Figs.~\ref{Fig2} and \ref{Fig3} are shown by  arrows.
The shaded triangle shows the SL phase.
For the 3D phase diagram and phase boundaries see SM \cite{supp}}.
\label{Fig1}
\vskip -0.5cm
\end{figure}

\emph{Model.}---%
The  general NN TL model \cite{SciRep,Chen3} with spin anisotropies constrained by the TL symmetries 
has both $XXZ$ and bond-dependent terms, ${\cal H}\!=\!{\cal H}_{XXZ} + {\cal H}_{\rm bd}$, 
\vskip -0.15cm \noindent
\begin{eqnarray}
&&{\cal H}_{XXZ}=J\sum_{\langle ij\rangle} 
\left(S^{x}_i S^{x}_j+S^{y}_i S^{y}_j+\Delta S^{z}_i S^{z}_j\right),\nonumber\\
\label{H}
&&{\cal H}_{\rm bd}=\sum_{\langle ij\rangle} 2J_{\pm\pm}
\left(\cos{\tilde{\varphi}_\alpha} \left[x,y\right]_{ij}-\sin{\tilde{\varphi}_\alpha} \left\{x,y\right\}_{ij}\right)\ \  \ \ \ \ \\
&&\phantom{{\cal H}_{\rm bd}\,\,\sum_{\langle ij\rangle}} +J_{z\pm}
\left(\cos{\tilde{\varphi}_\alpha} \left\{y,z\right\}_{ij} -\sin{\tilde{\varphi}_\alpha} \left\{x,z\right\}_{ij}\right),\nonumber
\end{eqnarray}
\vskip -0.35cm \noindent
where $0\!\leq\!\Delta\!\leq\!1$ for  layered systems, auxiliary phases are 
$\tilde{\varphi}_\alpha\!=\!\{0,-2\pi/3,2\pi/3\}$
for bonds along the primitive vectors ${\bm \delta}_\alpha$ in  Fig.~\ref{Fig1}, and  notations
$\left[a,b\right]_{ij}\!=\!S_i^aS_j^a\!-\!S_i^bS_j^b$ and 
$\left\{a,b\right\}_{ij}\!=\!S_i^a S_j^b\!+\!S_i^b S_j^a$ are used for brevity 
\cite{footnote1}.

\emph{Classical phase diagram.}---%
The $XXZ$ term in (\ref{H}) favors coplanar states: the well-known $120{\degree}$ state 
for $\Delta\!\leq\!1$ and $J\!>\!0$, and a ferromagnetic state for  $J\!<\!0$.
The terms in ${\cal H}_{\rm bd}$ in (\ref{H}) result in trends that are incompatible 
for different bonds \cite{supp}, leaving no continuous spin symmetries and selecting  ``stripe-{\bf x}'' and ``stripe-{\bf yz}''
as classical ground states \cite{SciRep,Chen3,Wang17}; see Fig.~\ref{Fig1}. 

In the  ``stripe-{\bf x}'' state, favored by $J_{\pm\pm}\!<\!0$,  
spins align along one bond in ferromagnetic ``stripes'' that order antiferromagnetically, see Fig.~\ref{Fig1}. 
This structure is only partially frustrated as  the $J_{\pm\pm}$ term is fully satisfied on the $x$-bond and 
half-satisfied on two other bonds \cite{Wang17,supp}. The ``stripe-{\bf yz}'' state benefits the
$J_{\pm\pm}\!>\!0$ and $J_{z\pm}$ terms in a similar manner.
Here spins in ferromagnetic stripes are perpendicular to the fully satisfied $x$-bond and are 
tilting out of the lattice plane with the angle dependent on the ratio $J_{z\pm}/J_{\pm\pm}$, 
reaching $\pi/4$ at $J_{z\pm}\!\rightarrow\!\infty$ \cite{Wang17}.
The boundaries between all  phases  in  Fig.~\ref{Fig1}
can be found analytically; see the Supplemental Material (SM), Ref.~\cite{supp}. 
The results are identical for $J_{z\pm}\!<\!0$ \cite{Chen3}.

In  Fig.~\ref{Fig1}, we present a two-dimensional (2D) cut of the classical 3D phase diagram of the NN TL 
model (\ref{H})  at $\Delta\!=\!1.0$ that shows all four phases discussed above. 
The full 3D phase diagram is a solid cylinder with the vertical axis $0\!\leq\!\Delta\!\leq\!1$
and the 2D cuts showing only quantitative 
changes vs $\Delta$; see the SM \cite{supp}.
The polar parametrization maps the 2D  parameter space onto a circle
and the choice of  numerical factors   
is to exaggerate the region $J_{\pm\pm}, J_{z\pm}\!\alt\!J$. 
In  Fig.~\ref{Fig1}, we also identify an area of the suspected SL phase discussed below.
It has a similarly limited extent along the $\Delta$-axis of the 3D phase diagram.

\begin{figure}[t]
\includegraphics[width=\linewidth]{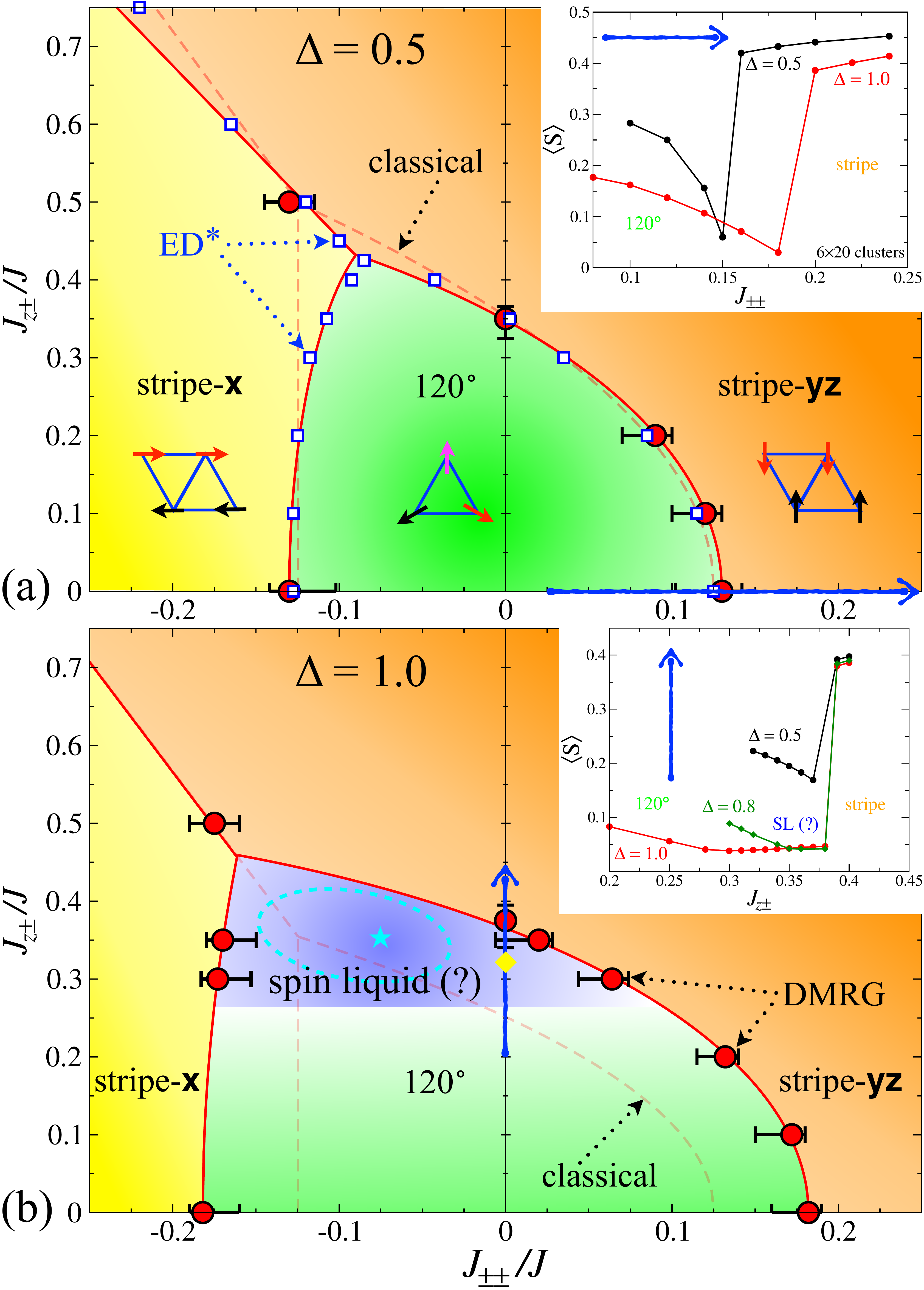}
\vskip -0.3cm
\caption{The 2D phase diagrams for $\Delta\!=\!0.5$ and $1.0$.
The circles are from the 1D DMRG scans, see text and 
Fig.~\ref{Fig3}, squares are ED results \cite{Wang17}, solid lines are guides to the eye,
dashed lines are classical phase boundaries, arrows show parameter cuts for the insets.
Insets show $\langle S\rangle$ vs $J_{\pm\pm}$ ($J_{z\pm}$) in units of $J$ 
from cylinders with fixed parameters \cite{footnote_sizes}.
Stars are parameters used in $1/L$ scaling in Fig.~\ref{Fig3}(c), see text.}
\label{Fig2}
\vskip -0.5cm
\end{figure}

\emph{DMRG results.}---%
To investigate the 3D phase diagram of the model (\ref{H}) by DMRG we use several complementary 
approaches. First is the long-cylinder 1D  ``scans,'' in which one of the parameters  
is varied along the length of the cylinder and   spin patterns provide a faithful 
visual extent of different phases that appear \cite{ZhuWhite,hexZhuWhite,us}. 
We use different boundary conditions and ranges of the varied 
parameter to exclude unwanted proximity effects \cite{supp}.
Second are the shorter cylinders \cite{footnote_sizes} with fixed parameters [``non-scans''] used as a probe 
for a sequence of points along   
the same 1D scans or at individual points of the phase diagram. 
Third is the $1/L$ scaling of the ordered moment 
using clusters with fixed aspect ratio \cite{white07}.
We also use measurements of the correlation lengths and intensity maps of the structure factor, ${\cal S}({\bf q})$ \cite{supp}.

\begin{figure}[t]
\includegraphics[width=\linewidth]{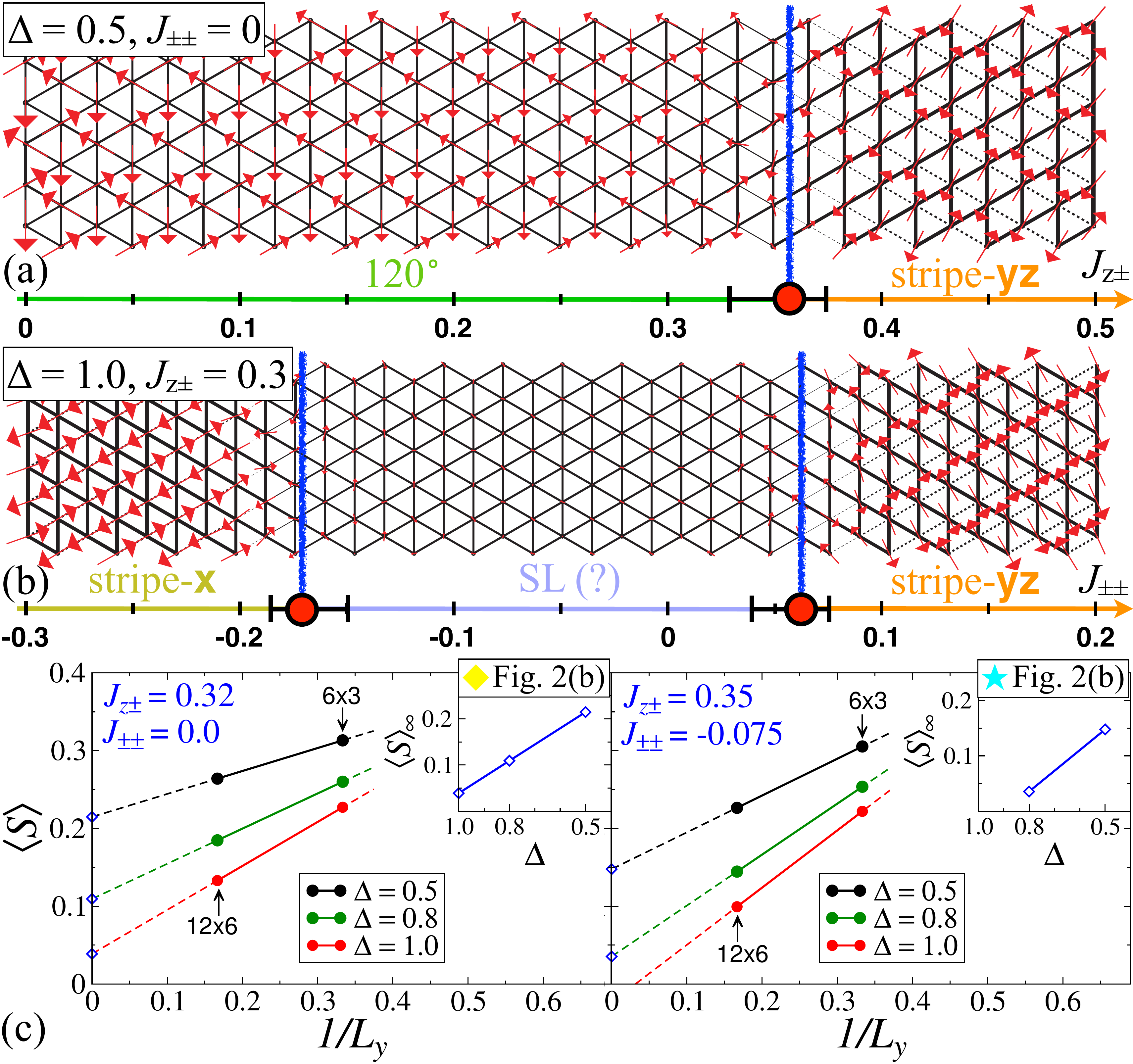}
\vskip -0.2cm
\caption{The 1D DMRG scans \cite{footnote_sizes} 
for (a) $\Delta\!=\!0.5$, $J_{\pm\pm}\!=\!0$ vs $J_{z\pm}$, 
and (b) $\Delta\!=\!1.0$, $J_{z\pm}\!=\!0.3$ vs $J_{\pm\pm}$ [$J_{\pm\pm}$, 
$J_{z\pm}$  in units of $J$]. The circles and lines show transitions, arrows 
are the in-plane projections of $\langle S\rangle$. 
(c) The $1/L$ scaling of $\langle S\rangle$ for parameters marked with stars 
in Fig.~\ref{Fig2}(b). Insets show  $\Delta$-dependencies of the extrapolated $\langle S\rangle_\infty$.
}
\label{Fig3}
\vskip -0.6cm
\end{figure}

Figs.~\ref{Fig2} and \ref{Fig3} present our key results; see the SM \cite{supp} for details.
In Fig.~\ref{Fig2} we show  2D phase diagrams for $\Delta\!=\!0.5$ and $1.0$, focusing on 
the region around the $120{\degree}$ phase.
The circles with error bars are transitions observed in the scans, such as the ones
shown in Figs.~\ref{Fig3}(a) and (b); squares in Fig.~\ref{Fig2}(a) are the exact-diagonalization 
(ED) results from Ref.~\cite{Wang17}; solid lines are guides to the eye.

Our scans for $\Delta\!=\!0.5$ in Fig.~\ref{Fig2}(a) and the ED are remarkably close,
both showing direct transitions between robust magnetic orders 
that are nearly coincident with the classical phase boundaries (dashed lines). 
The non-scan ordered moments,
shown in the insets of Fig.~\ref{Fig2} along two representative cuts,
support these findings.   

Fig.~\ref{Fig2}(b) summarizes our results 
for the isotropic limit of the $XXZ$ term (\ref{H}), 
$\Delta\!=\!1.0$. We find  an expansion of the $120{\degree}$ phase beyond its 
classical boundaries with the transitions from it to the stripe phases remaining direct 
for  $J_{z\pm}\!\alt\!0.25$ \cite{supp}; see   inset in Fig.~\ref{Fig2}(a). 
Here we find a possible  SL state in the $J_{z\pm}\!\simeq\![0.27,0.45]$ window;
see also Fig.~\ref{Fig3}(b) and SM \cite{supp}. 

The inset of Fig.~\ref{Fig2}(b) presents the non-scan ordered moment $\langle S\rangle$ 
along the $J_{z\pm}$ cut. 
It shows a kink-like feature at  $J_{z\pm}\!\approx\!0.28$ for $\Delta\!=\!1.0$. However, the  
intermediate phase from $J_{z\pm}\!\approx\!0.28$ to $0.38$ still exhibits a weak order. 
The same plot shows a similar feature for $\Delta\!=\!0.8$ from   
$J_{z\pm}\!\approx\!0.35$ to $0.38$ while
for $\Delta\!=\!0.5$ the transition is direct.
The scans for $\Delta\!=\!0.8$  show the SL-suspect region that is significantly smaller than for
$\Delta\!=\!1.0$ \cite{supp}.
We  note that all transitions to stripe states that we observe are first-order like.

Another test of the SL region is provided by the $1/L$ scaling of the ordered moment 
$\langle S\rangle$ \cite{white07} in Fig.~\ref{Fig3}(c)  
for  representative points indicated by the stars in Fig.~\ref{Fig2}(b). 
The insets of Fig.~\ref{Fig3}(c) show the $\Delta$-dependence of the 
extrapolated moment $\langle S\rangle_\infty$.
This analysis suggests an SL state in a significantly smaller 3D region than the long-cylinder scans.
Its approximate extent  
in $\Delta\!=\!1.0$ plane is shown in Fig.~\ref{Fig2}(b) by the dashed oval and it is limited along the $XXZ$ axis 
by $\Delta\!\agt\!0.9$.

This should be compared with the $J_1$--$J_2$ model, where different 
methods used here agree very closely on the extent of the SL region  \cite{ZhuWhite,us,supp}. 
This is not the case in the present study, 
suggesting that a weak and/or   more complicated form ordering \cite{multiQ} 
may persist in much of the suspected SL region.

\begin{figure}[t]
\includegraphics[width=\linewidth]{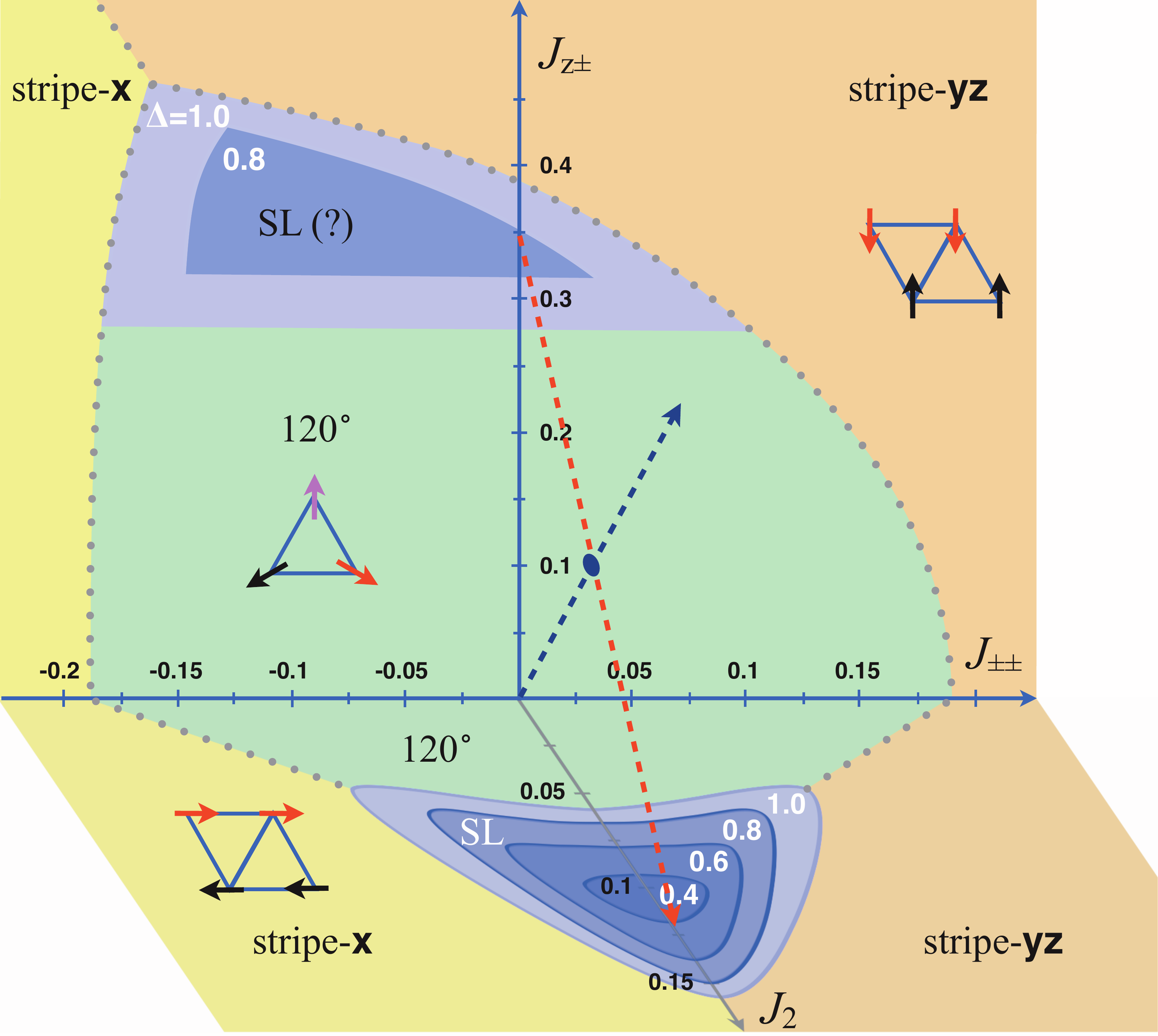}
\vskip -0.2cm
\caption{Topographic maps of the SL regions of the 3D phase diagrams 
of the NN ($J_2\!=\!0$) model (\ref{H})  [back panel] and 
of the $J_1$--$J_2$--$J_{\pm\pm}$ ($J_{z\pm}\!=\!0$) $XXZ$ model \cite{us} [lower panel]. $\Delta$-cuts are given in 0.2 increments and indicated. 
Dashed arrows show   1D DMRG cuts in Fig.~\ref{Fig5}.}
\label{Fig4}
\vskip -0.6cm
\end{figure}

We also verified that larger anisotropies  
enhance magnetic order and imply strongly gapped states,   
confirmed by the quasiclassical analysis \cite{supp}. 
One parameter choice that shows a nearly classical stripe order \cite{ModelB} was
originally advocated as a suitable SL for YMGO \cite{Chen4,Ruegg}.

We summarize the 3D quantum phase diagram for the model (\ref{H})  
in the back panel of Fig.~\ref{Fig4} as a topographic map. It retains all   classical phases of Fig.~\ref{Fig1} 
and acquires an SL region. The generous outline of the latter represents a distorted cone-like shape with 
the base at $\Delta=1.0$ and the widest dimensions $J_{z\pm}\!\simeq\![0.27,0.45]$ and 
$J_{\pm\pm}\!\simeq\![-0.17,0.1]$ at that base. 
The tip of the cone extends along the $XXZ$ axis down to 
$\Delta\!\agt\!0.7$. As is discussed above, the actual SL region may be significantly smaller.
We also note that the SL phase occurs within the $120{\degree}$ region and its maximal extent is achieved 
at the isotropic limit of the $XXZ$ term, questioning the importance of anisotropies for
its existence. 

\emph{$J_2$--extension.}---%
Some of the most reliable experiments in YMGO strongly suggest
that one should add a second-NN $J_2$-term to the NN model (\ref{H}) 
 \cite{MM,MM2}.
The isotropic $J_1$--$J_2$ model is also known to have
an SL state for a range of $J_2\!\approx\![0.06,0.16]J_1$ 
\cite{Bishop,Mishmash13,ZhuWhite,Yigbal,Kaneko,Sheng1,Sheng2,McCulloch,schwinger17}.
For both reasons, a minimal  modification of the NN model (\ref{H})
by the $XXZ$-only next-NN $J_2$-term suffices \cite{supp}.
 
Recently, we have investigated
the effect of the $XXZ$ and $J_{\pm\pm}$ anisotropic terms on the $J_1$--$J_2$ SL phase \cite{us}.
It survives down to $\Delta\!\approx\!0.3$ and is eliminated completely 
by $|J_{\pm\pm}|\!\approx\!0.1$.
In the bottom panel of Fig.~\ref{Fig4}, we present a topographic map of the SL state in this 
$XXZ$ $J_1$--$J_2$--$J_{\pm\pm}$ ($J_{z\pm}\!=\!0$) model 
using results from Ref.~\cite{us}. 

Fig.~\ref{Fig4} suggests a connection between the SL phases of the anisotropic 
model (\ref{H}) and of its isotropic $J_1$--$J_2$ counterpart. 
We verify this connection for $\Delta\!=\!1.0$ where  
the extent of both SL regions in Fig.~\ref{Fig4} is maximal. We use two long-cylinder DMRG  scans in the  
$J_{\pm\pm}\!=\!0$ plane shown in Fig.~\ref{Fig4} by  arrows. 
The first scan connects the anisotropic SL at $J_{z\pm}\!=\!0.35$ with 
the isotropic $J_1$--$J_2$ SL at $J_2\!=\!0.12$ (units of $J_1$).
The second scan starts at the origin ($J_1$-only model) and is used to confirm the existence of an  SL state between  
the $120{\degree}$ and the stripe phases along the direction tilted from the $J_2$ and $J_{z\pm}$ axes. 
For a different picture of these cuts, see  SM \cite{supp}.

\begin{figure}[t]
\includegraphics[width=\linewidth]{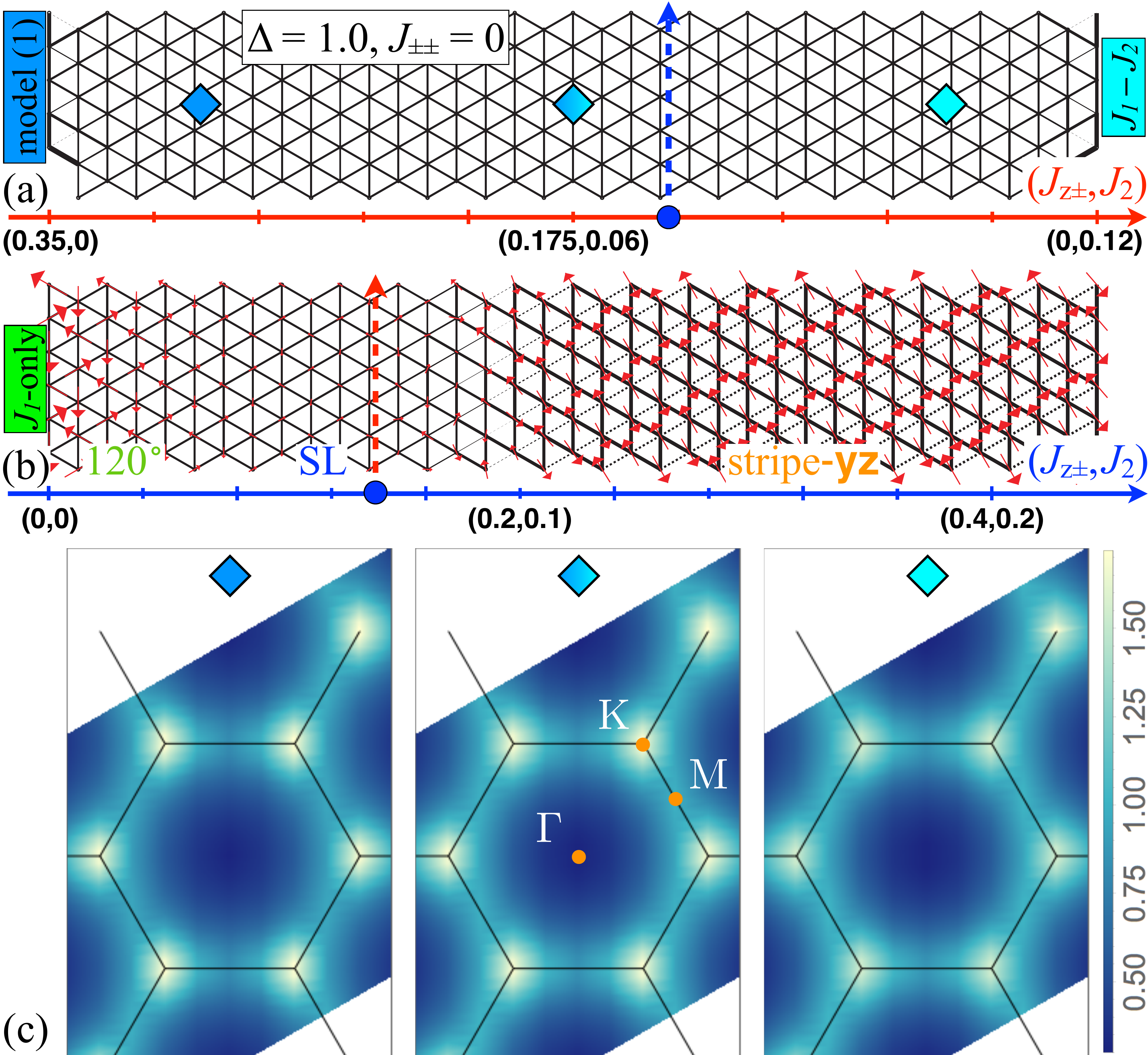}
\vskip -0.2cm
\caption{(a) and (b) DMRG  scans at $\Delta\!=\!1.0$  
as shown in Fig.~\ref{Fig4}. $J_{z\pm}$, $J_2$ are in units of $J_1$; the crossing point
of the scans in Fig.~\ref{Fig4} is shown by  blue dots and dashed arrows.
(c) ${\cal S}({\bf q})$ maps from three section of the 
cylinder in (a)  marked by the diamonds.}
\label{Fig5}
\vskip -0.4cm
\end{figure}

Figs.~\ref{Fig5}(a) and (b) show the real-space images of these scans; dashed arrows are marking 
their crossing point; both coordinates  $(J_{z\pm}, J_2)$ are indicated.
The first cylinder has open boundary condition and one site removed at each end  to suppress  
spinon localization at the edge \cite{ZhuWhite,supp}. The scan shows no 
indication of  magnetic, chiral, or valence-bond order 
\cite{supp} and no change of the SL state along the cylinder, the latter inferred 
from the thickness of the bonds that are proportional to the nearest-neighbor correlation 
$\langle {\bf S}_i{\bf S}_j\rangle$.   
The second scan, Fig.~\ref{Fig5}(b), shows a transition from the $120{\degree}$ to the stripe-${\bf yz}$ state
via an intermediate SL state, consistent with the first scan and also with the results in 
Fig.~\ref{Fig2}(b) and Fig.~\ref{Fig4} and in Refs.~\cite{us,supp} for the scans of the same nature along 
$J_{z\pm}$ and $J_2$ axes.

To infer the character of the SL states, we calculate 
the static structure factor, ${\cal S}({\bf q})$,
using correlations from the three section of the cylinder in Fig.~\ref{Fig5}(a) with centers of these 
sections marked by the diamonds \cite{footnoteSq0}.
The first section represents the region that is close to the limit of the original anisotropic model (\ref{H}), the third
section is close to the  isotropic $J_1$--$J_2$ SL, and the second is in between.
The results are shown in Fig.~\ref{Fig5}(c) where ${\cal S}({\bf q})$ is at $q_z\!=\!0$ \cite{footnoteSq}.
We have also calculated  ${\cal S}({\bf q})$ in non-scan cylinders   
for the limits of the scan in Fig.~\ref{Fig5}(a) as well as at other points within the SL regions in Fig.~\ref{Fig4} 
with quantitatively very similar results \cite{supp}.

The structure factors in Fig.~\ref{Fig5}(c)  are nearly identical, implying that the SLs  in
the anisotropic model (\ref{H}) and   
in the  isotropic $J_1$--$J_2$ model, as well as any SL state in between, are isomorphic to each other \cite{footnoteSq1}.
The correlations show a broadened peak at the $K$-points, the feature consistent with the 
$Z_2$ \cite{ZhuWhite,Sheng1,schwinger17}, U(1) Dirac \cite{Yigbal},  or Dirac-like 
\cite{Balents17,Mishmash13,Kaneko,Sheng2,McCulloch} SLs,  
but not with the spinon Fermi surface  SL state proposed for YMGO \cite{Chen1,Chen7}. 
This  suggests that an extrinsic mechanism is responsible 
for an SL-like response in this material.
The YMGO structure factor has maxima of intensity at the $M$-points, the feature 
readily obtainable from   stripe domains of mixed orientations, see \cite{supp},
 supporting the SL mimicry scenario of    Ref.~\cite{us}.

\emph{Summary.}---%
We have explored the 3D phase diagram of the anisotropic NN model on an ideal TL lattice (\ref{H})
using DMRG approaches. In agreement with  prior work, well-ordered states 
dominate most of it, showing reduced quantum fluctuations and strongly gapped states  for significant anisotropic interactions, also confirmed by the quasiclassical analysis. 
We have identified an SL region of the phase diagram and created its topographic map. This SL 
state occurs at the border between $120{\degree}$ and stripe phases, with its maximal extent 
reached at the isotropic limit of the $XXZ$ term.   
We have studied a four-dimensional extension of the phase diagram by the next-NN $J_2$-term
and have shown that it connects the newly found SL state  
to the well-known isotropic $J_1$--$J_2$  model. 
The spin-spin correlations are nearly identical everywhere between these two limits, 
 suggesting a complete isomorphism of the corresponding SL states. This also 
rules out the ``spinon metal'' SL state as a viable candidate for materials that realize anisotropic TL model.
We also note that given its relative volume of the phase diagram, realizing the newly discovered SL state 
in a real material  would require a  
parameter fine-tuning. 

\begin{acknowledgments}
\emph{Acknowledgments.}---%
We thank Prof. Xiaoqun Wang and Dr. Qiang Luo for sending their published data from Ref.~\cite{Wang17},
Itamar Kimchi for sharing his notes prior to publication and a discussion, and Cecile Repellin for sharing 
unpublished results and a useful conversation.
We are grateful to George Jackeli for a patient discussion and support and to Natalia Perkins for a useful technical tip. 
We are indebted to Martin Mourigal for numerous communications, explanations, 
indispensable comments, and unique insights.
This work was supported by the U.S. Department of Energy,
Office of Science, Basic Energy Sciences under Award No. DE-FG02-04ER46174 (P. A. M. and A. L. C.)
and by the NSF through grant DMR-1505406 (Z. Z. and S. R. W.).
\end{acknowledgments}
\vskip -0.3cm \noindent



\newpage
\onecolumngrid
\begin{center}
{\large\bf Topography of Spin Liquids on a Triangular Lattice:  Supplemental Material}\\ 
\vskip0.35cm
Zhenyue  Zhu,$^1$ P.  A. Maksimov,$^1$ Steven R. White,$^1$ and A. L. Chernyshev$^1$\\
\vskip0.15cm
{\it \small $^1$Department of Physics and Astronomy, University of California, Irvine, California
92697, USA}\\
{\small (Dated: February 9, 2018)}\\
\vskip 0.1cm \
\end{center}
\twocolumngrid

\setcounter{equation}{0}
\setcounter{figure}{0}

\subsection{Model and classical phase diagram}
\vskip -0.3cm

\begin{figure}[t]
\includegraphics[width=0.8\linewidth]{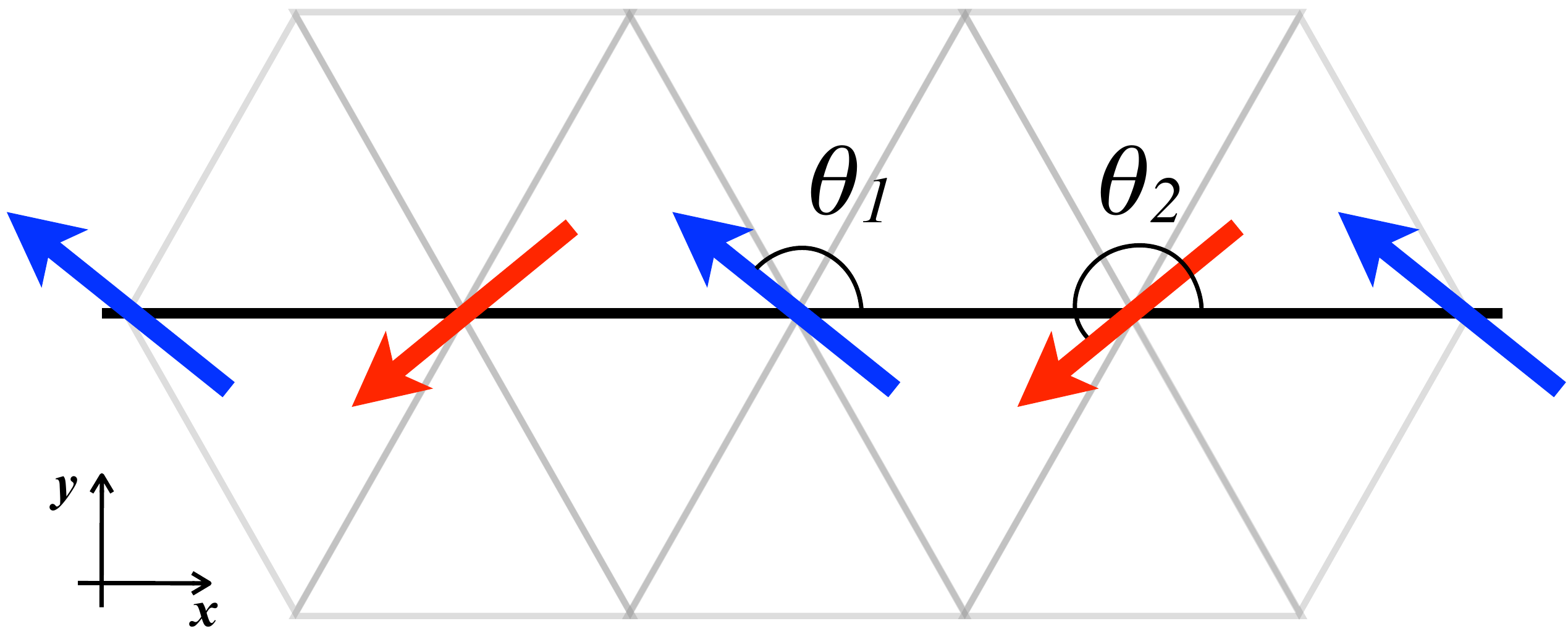}
\caption{An example of a counter-rotating spins configuration that minimizes 
classical energy of the $J_{\pm\pm}\!<\!0$ term along the $x$-axis bond with 
$\theta_2\!=\!-\theta_1$ (mod $2\pi$).}
\label{s_Fig1}
\vskip -0.5cm
\end{figure}

\emph{Model.}---%
The nearest-neighbor model on an ideal triangular lattice with spin anisotropies 
constrained by the lattice symmetries \cite{sChen1} can be written as a sum of the conventional
$XXZ$ and bond-dependent terms, 
\begin{eqnarray}
&&{\cal H}\!=\!{\cal H}_{XXZ}+{\cal H}_{\pm\pm}+{\cal H}_{z\pm},\nonumber\\
\label{sH}
&&{\cal H}_{XXZ}=J_1\sum_{\langle ij\rangle} 
\left(S^{x}_i S^{x}_j+S^{y}_i S^{y}_j+\Delta S^{z}_i S^{z}_j\right),\\
&&{\cal H}_{\pm\pm}=2J_{\pm\pm}\sum_{\langle ij\rangle} 
\left(\cos{\tilde{\varphi}_\alpha} \left(S_i^xS_j^x-S_i^yS_j^y\right)\right.\nonumber\\
\label{sHbd}
&&\phantom{{\cal H}_{\pm\pm}=J_{\pm\pm}\sum_{\langle ij\rangle}}
\left.-\sin{\tilde{\varphi}_\alpha} \left(S_i^x S_j^y+S_i^y S_j^x \right)\right),\ \ \\
&&{\cal H}_{z\pm}=J_{z\pm}\sum_{\langle ij\rangle} 
\left(\cos{\tilde{\varphi}_\alpha} \left(S_i^y S_j^z+S_i^z S_j^y \right)\right.\nonumber\\
&&\phantom{{\cal H}_{z\pm}=J_{z\pm}\sum_{\langle ij\rangle}}
\left.-\sin{\tilde{\varphi}_\alpha} \left(S_i^x S_j^z+S_i^z S_j^x \right)\right),\nonumber
\end{eqnarray}
where the $XXZ$ anisotropy  is taken $0\!\leq\!\Delta\!\leq\!1$ (typical for layered systems), 
and the ``$120{\degree}$ phase factors'' of the triangular-lattice bond directions  
along the primitive vectors ${\bm \delta}_\alpha$ in Fig.~\ref{s_Fig1}
are $\tilde{\varphi}_\alpha\!=\!\{0,-2\pi/3,2\pi/3\}$.

We note that we use a standard form of the $XXZ$ term in (\ref{sH}) (cf.  \cite{smultiQ,sChen1,sWang17} 
and other works).
To relate to the notations of these other works, our 
$J_1\!=\!2{\sf J_\pm}$ and anisotropy $\Delta\!=\!{\sf J_z}/2{\sf J_\pm}$.  
These works also use  ${\sf J_z}\!=\!\Delta J_1$ as a unity, while we use a more natural $J_1$ in that role.
Although our anisotropic (bond-dependent) terms are the same as elsewhere, we use a different operator form 
of them in (\ref{sHbd}) that is significantly more transparent as to what states these terms may be favoring.

\emph{Classical states.}---%
For the antiferromagnetic sign of $J_1$ and $\Delta\!\leq\!1$, the $XXZ$ term (\ref{sH}) 
favors the well-known $120{\degree}$ state with a plane that is coplanar with the plane of the lattice.
It favors a ferromagnetic state if  $J_1\!<\!0$.

The bond-dependent terms in (\ref{sHbd}) have conflicting trends along different bond directions. 
Consider the $J_{\pm\pm}$ term along ${\bm \delta}_1$ ($x$-axis bond, $\tilde{\varphi}_1\!=\!0$). 
Taking this bond direction in isolation, i.e., along  the $x$-axis  1D chain, this term
is given by $2J_{\pm\pm}\sum_{\langle ij\rangle}\left(S_i^xS_j^x-S_i^yS_j^y\right)$.
As one can see from a transformation $S_i^y\!\rightarrow\!-S_i^y$ at every second site of the chain,
this term preserves a U(1) symmetry in the $x$-$y$ plane (plane of the lattice).
The classical set of states that is favored by such a term is a two-sublattice configuration of  
counter-rotating spins, which minimizes
$J_{\pm\pm}\cos\left(\theta_1+\theta_2\right)$ at every bond, here
$\theta_1$ and $\theta_2$ are the in-plane angles of spins in the two sublattices, respectively.
Thus,  $\theta_2\!=\!\pi-\theta_1$  for  $J_{\pm\pm}\!>\!0$ and $\theta_2\!=\!-\theta_1$  for  $J_{\pm\pm}\!<\!0$;
see Fig.~\ref{s_Fig1}.
This is not unlike the trend observed in the ``stripyhoneycomb'' 
$\gamma$-Li$_2$IrO$_3$ \cite{sColdea_Ir}, another material with strong spin-orbit-generated spin 
anisotropies where the counter-rotating spiral ordering has been observed and linked to the large Kitaev terms. 

The situation is similar for the $J_{z\pm}$ term along the $x$-axis bond, 
except that the plane of the counter-rotation of spins is $y$-$z$, i.e., spins are perpendicular to the optimized 
bond and are not contained to the plane of the lattice. The angle of the counter-rotation is also different,
as it is minimizing $J_{\pm\pm}\sin\left(\theta_1+\theta_2\right)$, so
$\theta_2\!=\!\pm\pi/2-\theta_1$ .

It is clear, however, that these 1D trends along different bond directions are
in conflict with each other. Consideration of the $J_{\pm\pm}$ term  
along the other bond directions ${\bm \delta}_\alpha$ 
gives the energy minimized by different angles of counter-rotation: 
$\theta_2\!=\!\pi-\theta_1-\tilde{\varphi}_\alpha$  for  $J_{\pm\pm}\!>\!0$ and 
$\theta_2\!=\!-\theta_1-\tilde{\varphi}_\alpha$  for  $J_{\pm\pm}\!<\!0$.
For the $J_{z\pm}$ term, the planes of spin counter-rotation are perpendicular to the corresponding bonds. 
That is, the angles (or planes in the $J_{z\pm}$ case) of the counter-rotated spins depend on the 
bond direction itself. Also, the triangular-lattice geometry is frustrating for the two-sublattice arrangements.

Therefore, even in the absence of the other competing terms,  because of the conflicting preferences on 
different bonds of the lattice, $J_{\pm\pm}$ and $J_{z\pm}$  interactions break continuous symmetries. 
We would like to stress that this mechanism of symmetry breaking 
is different from that of, e.g., Ising-like anisotropies, which explicitly break spin rotational symmetry on any bond.

\begin{figure}[t]
\includegraphics[width=0.85\linewidth]{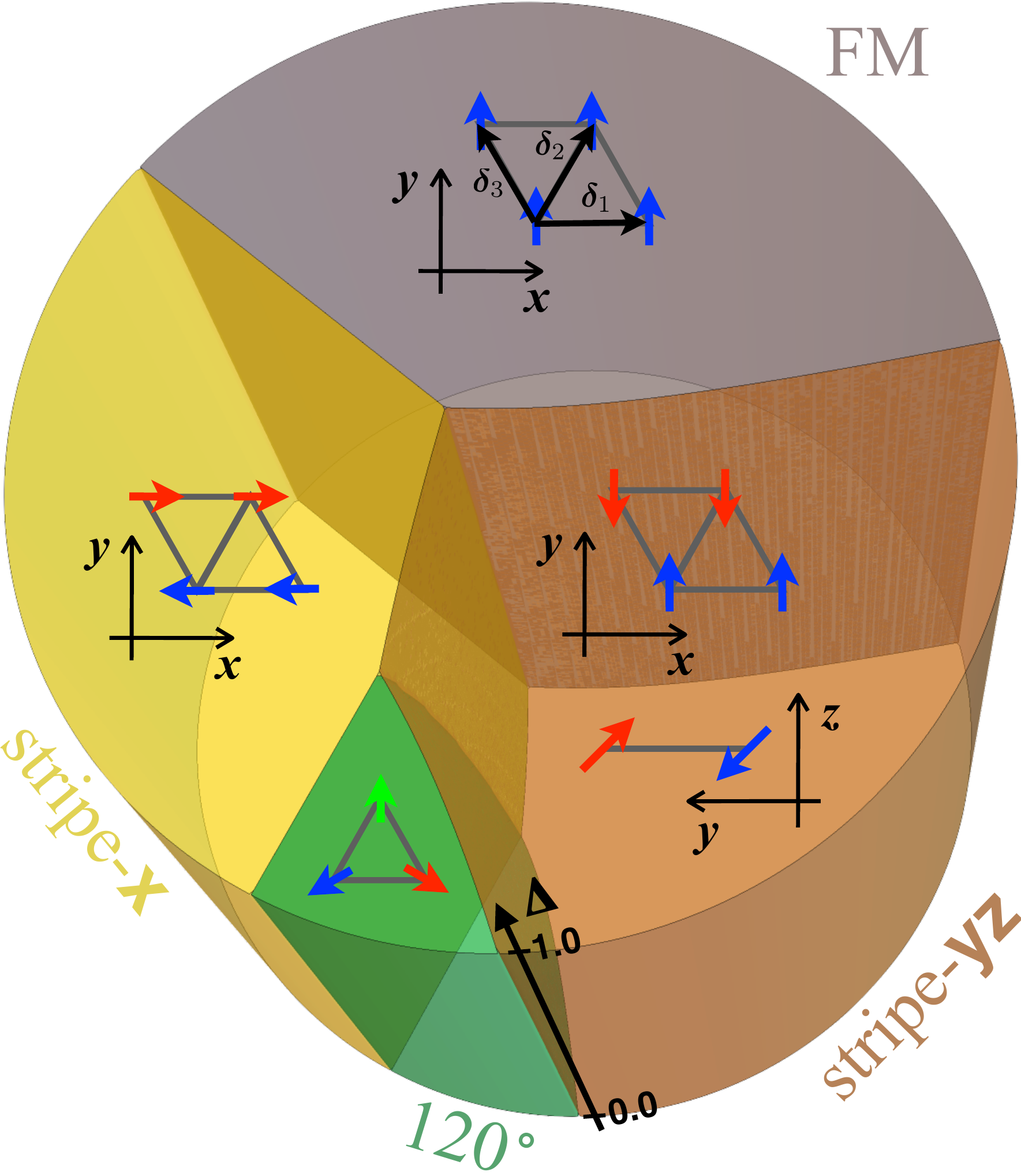}
\caption{The 3D classical phase diagram of the model (\ref{sH}), (\ref{sHbd}) 
represented using cylindrical parametrization of the parameters (see the text). 
Sketches of the phases together with primitive vectors of the lattice are shown and 
phases are discussed in the text in detail. See Fig.~1 of the main text for a 2D cut of this phase diagram at
$\Delta\!=\!1.0$ in more details.}
\label{s_Fig2}
\vskip -0.4cm
\end{figure}

As a result, only a discrete set of states can be favored by  $J_{\pm\pm}$ and $J_{z\pm}$  as 
the classical ground states, and collinear (``stripe'') phases of two types, 
referred to as ``stripe-{\bf x}'' and ``stripe-{\bf yz}'', are selected as such;
see Fig.~\ref{s_Fig2} and also Refs.~\cite{smultiQ,sWang17}. 
Thus, at the first glance, the above consideration of the counter-rotated spin 
configurations is completely irrelevant. However, this is not so, as, in fact, it explains
why stripes, and the stripes of these particular types, are selected.   

\emph{Stripe states.}---%
Consider the  ``stripe-{\bf x}'' state, favored by the $J_{\pm\pm}\!<\!0$ and unaffected by the $J_{z\pm}$ term.
Spins co-align with the $x$-axis bond and with each other along the same bond direction in a ferromagnetic 
chain structure (``stripe''). The ferromagnetic direction of the stripe alternates from one $x$-row to the other;
see Fig.~\ref{s_Fig2}. 
Needless to say, the ferromagnetic arrangements co-aligned with the $x$-bond are the two out of four 
collinear states from the counter-rotating manifold of states favored by $J_{\pm\pm}\!<\!0$
as far as the $x$-direction is concerned,  $\theta_2\!=\!-\theta_1\!=\!\pi$ and
$\theta_2\!=\!-\theta_1\!=\!0$; see Fig.~\ref{s_Fig1}. 
Therefore, classically, the $J_{\pm\pm}$ term is fully satisfied on the $x$-bond for the ``stripe-{\bf x}'' arrangement 
and the other two bonds are half-satisfied and, thus, are only partially frustrated  \cite{s_us}. 
The other two collinear states are antiferromagnetic $x$-chains
with the moments normal to the $x$-bond, $\theta_2\!=\!-\theta_1\!=\!\pm\pi/2$. They are more 
frustrated and are higher in energy.

In a sense, the resultant stripe phases are a result of locking together of the counter-rotating
preferences of the bond-dependent terms for different directions in a partially-frustrated fashion.
  
For the  $J_{\pm\pm}\!>\!0$ term,  selection of the ``stripe-{\bf yz}'' state is identical to the one above
with the only difference that the ferromagnetic moments now point perpendicular to the $x$-bond, 
along the $y$-axis, $\theta_2\!=\!\pi-\theta_1\!=\!\pi/2$, but are still in the plane of the lattice.

The $J_{z\pm}$ term also favors the ferromagnetic $x$-chain with  spins normal to the $x$-bond,
but with a tilt out of plane toward the $z$-axis, hence the name ``stripe-{\bf yz}''. 
The direction of ferromagnetic moment of stripes 
alternates from one $x$-row to the other; see Fig.~\ref{s_Fig2} and Ref.~\cite{sWang17}. If only the
$J_{z\pm}$ term is present, the tilt angle is $\pi/4$.
When both $J_{\pm\pm}\!>\!0$ and $J_{z\pm}$ terms are present, they both benefit from the ``stripe-{\bf yz}''
order with $x$-bond fully satisfied and the tilt angle interpolating 
between $0$ and $\pi/4$ depending on the ratio $J_{z\pm}/J_{\pm\pm}$.

\emph{Classical phase diagram.}---%
In Fig.~\ref{s_Fig1}, we present the classical 3D phase diagram of the nearest-neighbor triangular-lattice model
(\ref{sH}), (\ref{sHbd}) with all four phases discussed above. Fig.~1 of the main text contains its $\Delta\!=\!1.0$ cut.
We have used the cylindrical parametrization of the parameter space 
with the vertical axis mapped on $0\!\leq\!\Delta\!\leq\!1$, $J_{z\pm}$ as the radial, and 
$J_1$ and $J_{\pm\pm}$ as the polar variables, so that each horizontal cut represents 
the entire two-dimensional space of the model at fixed $\Delta$:
\begin{eqnarray}
\label{s_param}
\left(J_1,2J_{\pm\pm},5J_{z\pm}\right)=\left(-r\sin\varphi,r\cos\varphi,\sqrt{1-r^2}\right),
\end{eqnarray}
such that $\sqrt{J_1^2+(2J_{\pm\pm})^2+(5J_{z\pm})^2}\!=\!1$.
The particular choice of numerical coefficients in this parametrization
is to greatly exaggerate the region where all parameters are of the same order 
$J_{\pm\pm}, J_{z\pm}\!\alt\!J_1$. 

One can see that the only noticeable change vs $\Delta$ from the $XY$ ($\Delta\!=\!0$) 
to the Heisenberg  ($\Delta\!=\!1.0$)  limit of the $XXZ$ part of the model is a slight narrowing of the 
$120{\degree}$ phase in the $J_{z\pm}$ direction.  
The boundaries between all the phases shown in  Fig.~\ref{s_Fig1}
can be found analytically \cite{sWang17}. 
We reproduce them here with the additional boundaries for the ferromagnetic region:
\begin{align}
&\mbox{s-{\bf x} $-$ 120{\degree}}&&:\quad \eta=-\frac18 ,\nonumber\\
&\mbox{s-{\bf yz} $-$ 120{\degree}}&&:\quad \zeta=\sqrt{\left(\frac18-\eta\right)\left(\frac32-\Delta\right)} ,\nonumber\\
\label{s_phase_b}
&\mbox{s-{\bf x} $-$ s-{\bf yz}}&&:\quad \zeta=\sqrt{-2\eta\left(1-\Delta-4\eta\right)} ,\\
&\mbox{s-{\bf x} $-$ FM}&&:\quad \eta=1 \quad (J_1,\, J_{\pm\pm} < 0),\nonumber\\
&\mbox{s-{\bf yz} $-$ FM}&&:\quad \zeta=\sqrt{\left(3+\Delta\right)\left(1+\eta\right)} \quad (J_1 < 0) ,\nonumber
\end{align}
where the first column indicates the phases on the two sides of the transition line 
and the second the equation for that line. Dimensionless parameters are 
$\eta\!=\!J_{\pm\pm}/J_1$ and $\zeta\!=\!J_{z\pm}/J_1$ and abbreviations ``s-{\bf x}'' and  ``s-{\bf yz}'' for 
the ``stripe-{\bf x}'' and ``stripe-{\bf yz}'' were used. 

We note that more complicated, multi-${\bf Q}$ ordered states were suggested to occur 
near the $120{\degree}$ phase boundaries \cite{smultiQ}. 
We have also observed  indications of them in instabilities of magnon branches at  incommensurate wavevectors
in our spin-wave  calculations \cite{s_us}. 
Since corresponding regions appear to be very narrow, we do not elaborate on them any further.

\begin{figure*}[t]
\includegraphics[width=1\linewidth]{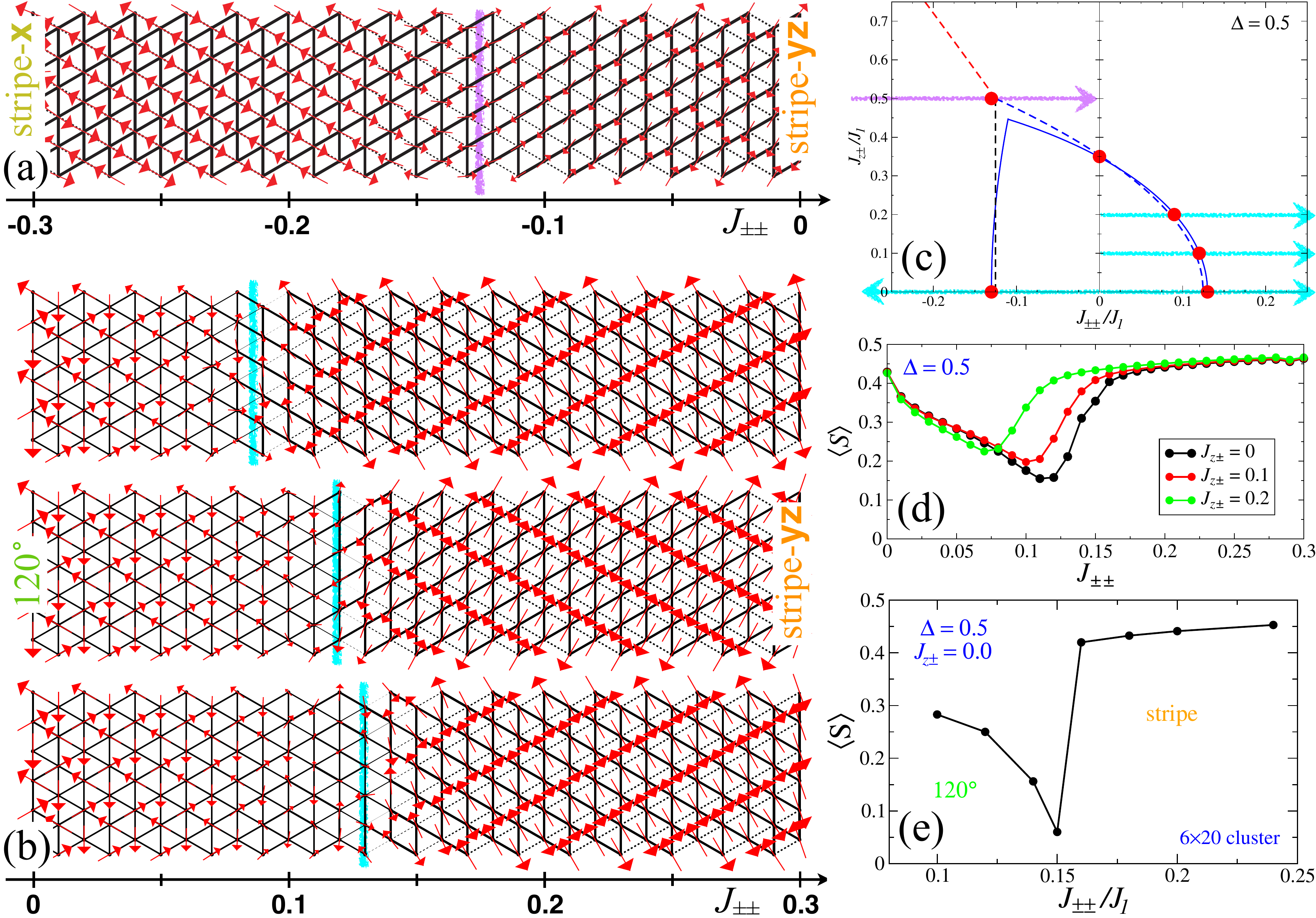}
\vskip -0.1cm
\caption{Results of the 1D long-cylinder DMRG scans for $\langle S\rangle$ vs $J_{\pm\pm}$ 
for various $J_{z\pm}$ and $\Delta\!=\!0.5$. (a) and (b) The real-space images of cylinders with arrows 
representing the in-plane $\langle S\rangle_{xy}$; vertical lines are inferred transition points.
(c) The directions of the cuts, transition points (circles), and classical (dashed lines) and 
quantum (solid lines) transition boundaries. (d) $\langle S\rangle$ vs $J_{\pm\pm}$ 
profiles for (b). (e) Is the same using fixed parameter, $6\!\times \!20$ cylinders, for 
$J_{z\pm}\!=\!0.0$ cut vs $J_{\pm\pm}$.}
\label{sFig3D05}
\vskip -0.3cm
\end{figure*}

\subsection{Details of the DMRG calculations}

For the DMRG calculations in the $6\!\times\!30$ and $6\!\times\!36$ cylinders, 
we perform 20 sweeps and typically keep up to 
$m\!=\!1600$ states with truncation error less than $10^{-5}$
and in some cases up to $m\!=\!2000$ states and 24 sweeps  depending on the complexity of the Hamiltonian. 
For the $6\!\times\!12$ and $6\!\times \!20$ cylinders, 
we typically perform 24 sweeps and keep up to $m\!=\!2000$ states with truncation errors less
than $10^{-6}$.
In the real-space images of cylinders, the size of the arrows represent the measurement of the projection
of  local spin in the $xy$ plane. The width of the bond on the lattice represents the 
nearest-neighbor spin-spin correlation, with ferromagnetic correlation shown as dashed and antiferromagnetic ones
as solid lines.

\begin{figure*}[t]
\includegraphics[width=1\linewidth]{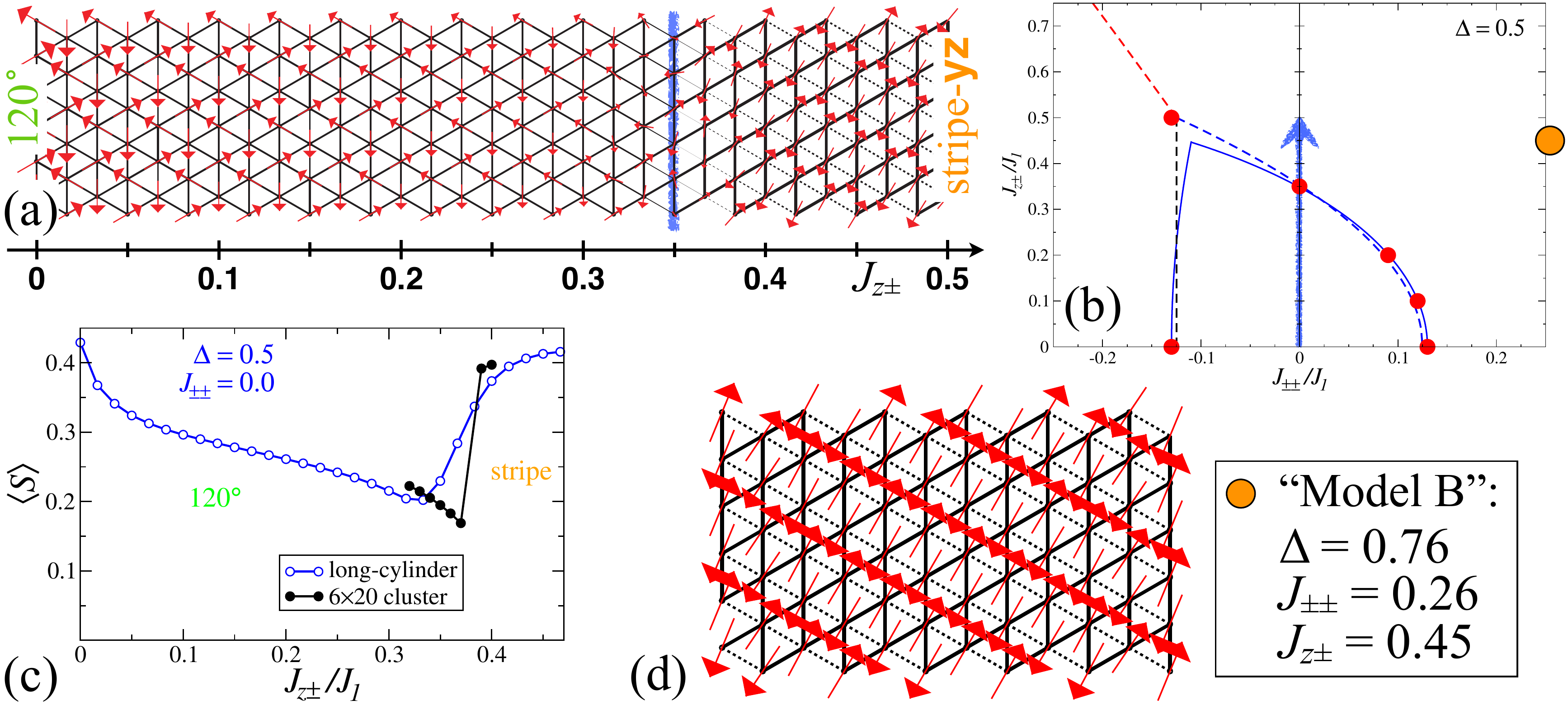}
\vskip -0.1cm
\caption{(a) Same as in Fig.~\ref{sFig3D05}(a) and (b) along the $J_{z\pm}$-axis as shown in (b).
(c) The  $\langle S\rangle$ vs $J_{z\pm}$  profiles from (a) and from $6\!\times \!20$ cylinders.
(d) The ground state for the ``Model B'' parameters (shown and indicated in (b) by the orange circle) in 
$6\times 12$ DMRG cylinder. The ordered moment is $\langle S\rangle\!\approx\!0.4694$.}
\label{sFig4D05}
\vskip -0.3cm
\end{figure*}

\emph{Phase diagram, $\Delta\!=\!0.5$.}---%
In Figs.~\ref{sFig3D05} and \ref{sFig4D05}  we provide a detailed 
summary of all our DMRG data for $\Delta\!=\!0.5$.
In Figs.~\ref{sFig3D05}(a) and (b), we show the real-space images of the  long-cylinder DMRG ``scans''
along the cuts indicated in Fig.~\ref{sFig3D05}(c), all varying $J_{\pm\pm}$ along their length and 
having all other parameters fixed. $J_{z\pm}$ is 0.5 for Fig.~\ref{sFig3D05}(a) scan and is 
0.2, 0.1, and 0.0, top to bottom, in Fig.~\ref{sFig3D05}(b), all in units of $J_1$ (\ref{sH}).   
The largest arrow in the cylinder images corresponds to the expectation  value of the in-plane
$\langle S\rangle_{xy}\!\approx\!0.43$ to $0.46$.
The results  at $J_{z\pm}\!=\!0.0$ are symmetric with respect to 
$J_{\pm\pm}\!\rightarrow\!-J_{\pm\pm}$ and ``stripe-{\bf yz}''$\rightarrow$``stripe-{\bf x}'', as indicated in 
 Fig.~\ref{sFig3D05}(c).

The  scan in Fig.~\ref{sFig3D05}(a) 
shows a direct transition between the ``stripe-{\bf x}'' state with spins along  the bonds
and the ``stripe-{\bf yz}'' state with spins perpendicular to the  bond and  tilted away from the $xy$ plane.
The scans in Fig.~\ref{sFig3D05}(b) show a direct transition between the $120{\degree}$ state 
and the ``stripe-{\bf yz}'' state. Profiles of $\langle S\rangle$ vs $J_{\pm\pm}$ in Fig.~\ref{sFig3D05}(d)
from these long-cylinder scans show no indication of a  magnetically disordered state.
Fig.~\ref{sFig3D05}(e) provides a point-by-point ``scan'' of the phase diagram using $6\!\times \!20$ cylinders
 with fixed parameters for  $J_{z\pm}\!=\!0.0$
vs $J_{\pm\pm}$ in a  range  $0.1$ to $0.24$; here  $\langle S\rangle$  is measured at the center 
of such cylinders. These results also suggest a direct transition 
between the $120{\degree}$ and  ``stripe-{\bf yz}'' states. 

In Fig.~\ref{sFig4D05}(a), we show the long-cylinder scan along the $J_{z\pm}$-axis at
$J_{\pm\pm}\!=\!0$, indicated in Fig.~\ref{sFig4D05}(b). It also demonstrates a clear direct transition 
between the $120{\degree}$ and  ``stripe-{\bf yz}'' states. Fig.~\ref{sFig4D05}(c) combines
$\langle S\rangle$ vs $J_{z\pm}$ profiles from this scan with the one from $6\!\times \!20$ cylinders
with fixed parameters, both unequivocally ruling out potential intermediate states.

\emph{Large anisotropies.}---%
We also make a brief note on the case of large anisotropic interactions in (\ref{sHbd}).

First is the case of a parameter choice that was advocated in Ref.~\cite{sChen3} as the prime 
candidate for a spin-liquid (SL) state  and was subsequently used in Ref.~\cite{sRuegg} with the same 
attitude. Following \cite{sRuegg}, we refer to this set of parameters as  the ``Model B'':
$\Delta\!=\!0.76$, $J_{\pm\pm}\!=\!0.26J$, $J_{z\pm}\!=\!0.45J$ (${\sf J_\pm}\!=\!0.66{\sf J_z}$, 
${\sf J_{\pm\pm}}\!=\!0.34{\sf J_z}$, ${\sf J_{z\pm}}\!=\!0.6{\sf J_z}$ in the notations of Refs.~\cite{sChen3,sRuegg}).
The justification of this choice was from the classical, high-temperature 
simulation, referred to as the self-consistent Gaussian appoximation, of the static structure 
factor, ${\cal S}({\bf q})$, yielding results that were seemingly in agreement with the experimental data.
While such an agreement is spurious because of an inconspicuous omission of the off-diagonal 
components of ${\cal S}({\bf q})$ in the calculations \cite{sMMnote},
we provide here the DMRG results for  Model B.  Although the $XXZ$ anisotropy 
is somewhat higher than in Fig.~\ref{sFig4D05}(b), we indicate the Model B $J_{\pm\pm}-J_{z\pm}$ 
coordinates by an orange circle. It is clearly deep in the stripe-{\bf yz} state, which is 
confirmed by the DMRG calculation in $6\!\times\! 12$  cylinder shown in Fig.~\ref{sFig4D05}(d).
The ordered moment is $\langle S\rangle\!\approx \! 0.4694$, which is also confirmed by  
spin-wave theory (SWT) calculations that yield  $\langle S\rangle\!\approx \! 0.4763$.
Both results indicate strongly  suppressed quantum fluctuations due to anisotropic terms.

Second case are the large values of $J_{z\pm}$. We choose $\Delta\!=\!1.0$ and $J_{\pm\pm}\!=\!0$
for simplicity. We have tested several $J_{z\pm}\!=\!4.0$, 8.0, and $\infty$.
DMRG yields the ground states that are virtually identical to the one in Fig.~\ref{sFig4D05}(d), i.e., they are
all nearly classical stripe-{\bf yz} states, with the out-of plane tilt angle changing as a function of $J_{z\pm}$ and 
reaching $\pi/4$ at $J_{z\pm}\!=\!\infty$, in agreement with the classical discussion above.
There is a slightly curious observation concerning the behavior of the almost classical ordered moments.
The fluctuations are slightly higher at $J_{z\pm}\!=\!\infty$ than at intermediate values $J_{z\pm}$, 
the behavior also replicated by the SWT:
$\langle S\rangle\!\approx \! 0.4886 (0.4914)$ [$J_{z\pm}\!=\!4.0$],
$\langle S\rangle\!\approx \! 0.4857 (0.4903)$ [$J_{z\pm}\!=\!8.0$], and 
$\langle S\rangle\!\approx \! 0.4795 (0.4869)$ [$J_{z\pm}\!=\!\infty$] by DMRG (SWT).

No indication of a massive degeneracy due to large anisotropic interactions has been identified.

\begin{figure*}[t]
\includegraphics[width=1\linewidth]{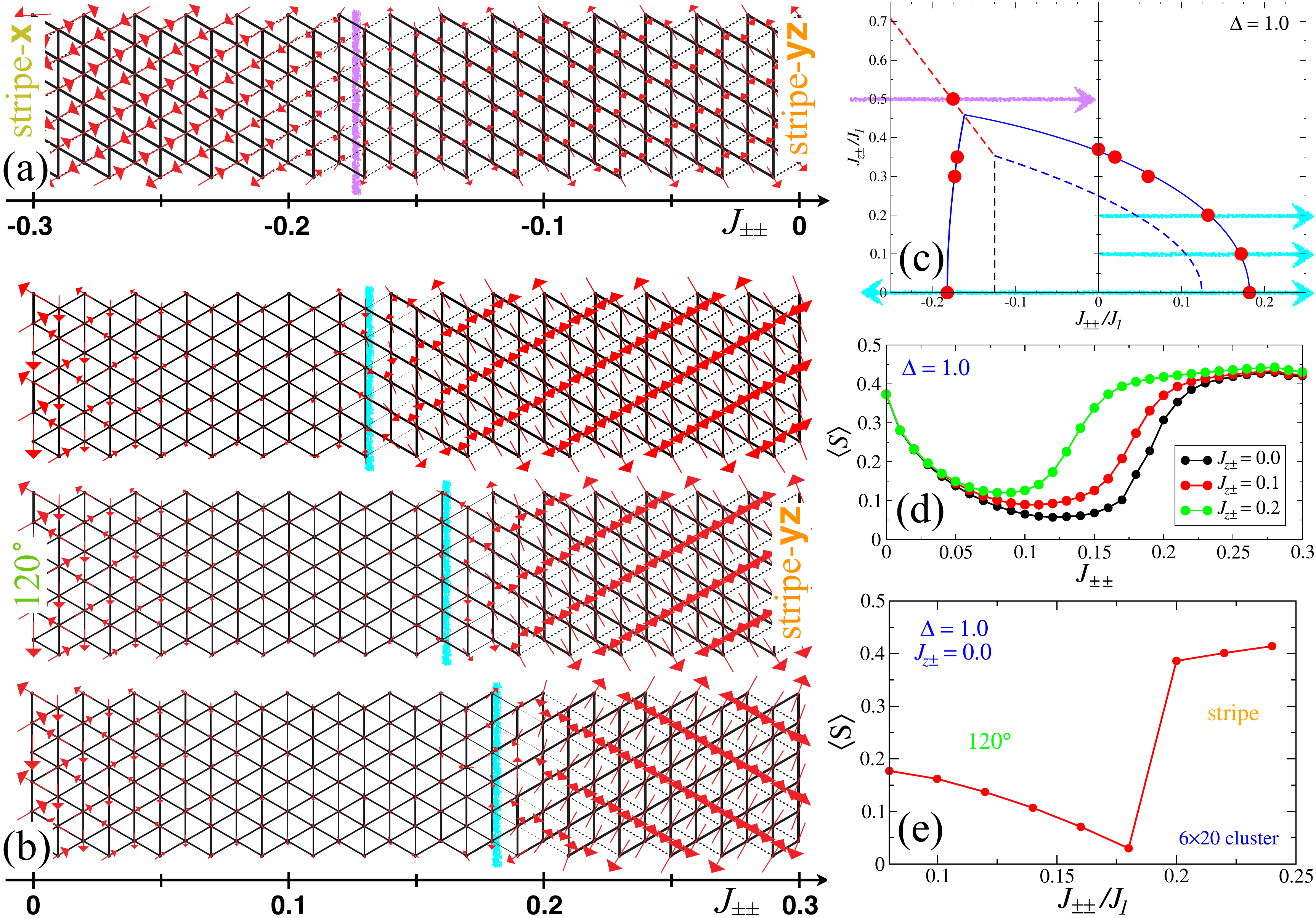}
\vskip -0.1cm
\caption{Results of the 1D long-cylinder DMRG scans for $\langle S\rangle$ vs $J_{\pm\pm}$ 
for various $J_{z\pm}$ and $\Delta\!=\!1.0$. (a) and (b) The real-space images of cylinders with arrows 
representing the in-plane $\langle S\rangle_{xy}$; vertical lines are inferred transition points.
(c) The directions of the cuts, transition points (circles), and classical (dashed lines) and 
quantum (solid lines) transition boundaries. (d) $\langle S\rangle$ vs $J_{\pm\pm}$ 
profiles for (b). (e) Is the same using fixed parameter, $6\!\times \!20$ cylinders, for 
$J_{z\pm}\!=\!0.0$ cut vs $J_{\pm\pm}$.}
\label{sFig5D1}
\vskip -0.3cm
\end{figure*}

\begin{figure*}[t]
\includegraphics[width=1\linewidth]{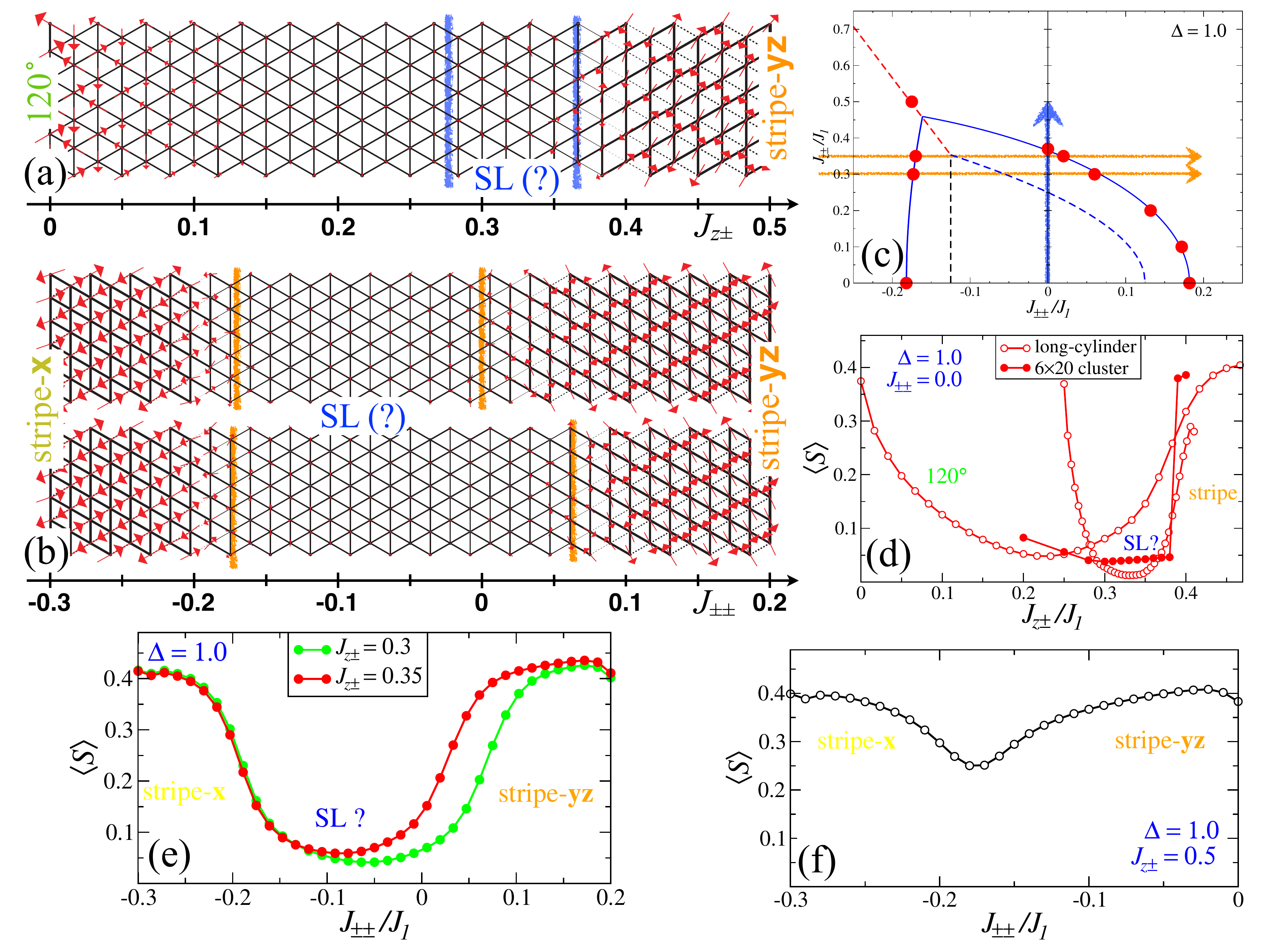}
\caption{(a) and (b) Same as in Fig.~\ref{sFig5D1}(a) and (b) along the $J_{z\pm}$-axis
and  parallel to the $J_{\pm\pm}$-axis at $J_{z\pm}\!=\!0.3$ and 0.35 as is shown in (c).
(d) The  $\langle S\rangle$ vs $J_{z\pm}$  profiles from (a), a long cylinder with smaller
$J_{z\pm}$ range, and from $6\!\times \!20$ cylinders. 
(e) The  $\langle S\rangle$ vs $J_{\pm\pm}$  profiles from the scans in (b). 
(f) The  $\langle S\rangle$ vs $J_{\pm\pm}$  profile from the scan in Fig.~\ref{sFig5D1}(a).}
\label{sFig6D1}
\vskip -0.3cm
\end{figure*}

\emph{Phase diagram, $\Delta\!=\!1.0$.}---%
Our summary of the DMRG data for $\Delta\!=\!1.0$ in Figs.~\ref{sFig5D1} and \ref{sFig6D1}
is structured similarly to the one for  $\Delta\!=\!0.5$ above. The directions of the  long-cylinder DMRG scans
for Figs.~\ref{sFig5D1}(a) and (b) are indicated in Fig.~\ref{sFig5D1}(c).
As before, the transition at a large $J_{z\pm}\!=\!0.5$ in Fig.~\ref{sFig5D1}(a) 
is a direct one  between the stripe-{\bf x} and the stripe-{\bf yz} states
and Fig.~\ref{sFig6D1}(f) demonstrates that by the profile of $\langle S\rangle$ throughout this transition.
Although this is harder to appreciate visually in the real-space images of cylinders 
in Fig.~\ref{sFig5D1}(b), they still show direct transitions between the $120{\degree}$  
and the stripe-{\bf yz} state for $J_{z\pm}=$0.2, 0.1, and 0.0, with the 
profiles of $\langle S\rangle$ vs $J_{\pm\pm}$ from these long-cylinder scans shown in Fig.~\ref{sFig5D1}(d).
This is verified by the $6\!\times \!20$ cylinders ``scan'' with fixed parameters for  $J_{z\pm}\!=\!0.0$ in
Fig.~\ref{sFig5D1}(e). Although the $120{\degree}$ order parameter is reduced, this analysis also supports 
a direct transition without an intermediate state.

Another observation from these data is that the $120{\degree}$ state has expanded beyond its classical 
phase boundaries at the cost of the stripe states, as discussed in the main text. This can be
expected as the excitations in the $120{\degree}$ state remain gapless despite anisotropic terms
and, therefore, allow for more quantum fluctuations to lower its energy 
as opposed to the stripe phases that are all gapped.

In Figs.~\ref{sFig6D1}(a) and (b) we show the long-cylinder scans along the $J_{z\pm}$-axis at
$J_{\pm\pm}\!=\!0$ and two additional scans  that are parallel to the $J_{\pm\pm}$-axis 
at $J_{z\pm}\!=\!0.3$ and 0.35, respectively, with the cuts indicated in Fig.~\ref{sFig6D1}(b). 
At the first glance, the $\langle S\rangle$ vs $J_{z\pm}$ profile from the scan in Fig.~\ref{sFig6D1}(a) is not too different
from, e.g.,  one of the profiles in  Fig.~\ref{sFig5D1}(d). On a closer inspection 
that uses a smaller range of $J_{z\pm}$ and thus a smaller gradient of $J_{z\pm}$, one can observe 
an indication of the magnetically disordered (SL) state between the $120{\degree}$  and the stripe-{\bf yz} states;
see Fig.~\ref{sFig6D1}(d). The same Fig.~\ref{sFig6D1}(d) shows the results of the 
$6\!\times \!20$ cylinders ``scan'' of the same area with fixed parameters. It shows a distinct ``kink'' in 
$\langle S\rangle$ at about $J_{z\pm}\!\approx\!0.28$, indicative of a transition.
One can see, however, that the intermediate phase from  $J_{z\pm}\!\approx\!0.28$ to 0.38 
still exhibits a weak order. This, combined with the $1/L$-scaling analysis below may suggest 
that a weak and/or a more complicated form of ordering referred to as the multi-${\bf Q}$ states
\cite{smultiQ} cannot be fully ruled out for that region. We, therefore, mark it as a ``suspected'' spin-liquid, or
``SL ?'' region.

The two other long-cylinder DMRG scans in Fig.~\ref{sFig6D1}(b) 
span the range of $J_{\pm\pm}$ that connects the two stripe phases.
The profiles of $\langle S\rangle$ vs $J_{\pm\pm}$ from these scans shown in Fig.~\ref{sFig6D1}(e) 
show a clear trend to a much suppressed order between the two stripe phases, which is
clearly distinct from the scans for the smaller $J_{z\pm}$ and for $\Delta\!=\!0.5$ case.
Fig.~\ref{sFig6D1}(e) indicates that the suspected SL region covers the entire area between the stripe-{\bf x} 
and stripe-{\bf yz} phases
above $J_{z\pm}\!\approx\!0.28$ and up to the tricritical point at  
$[J_{\pm\pm}\!\approx\!-0.17, J_{z\pm}\!\approx\!0.45]$. 

\begin{figure*}[t]
\includegraphics[width=1\linewidth]{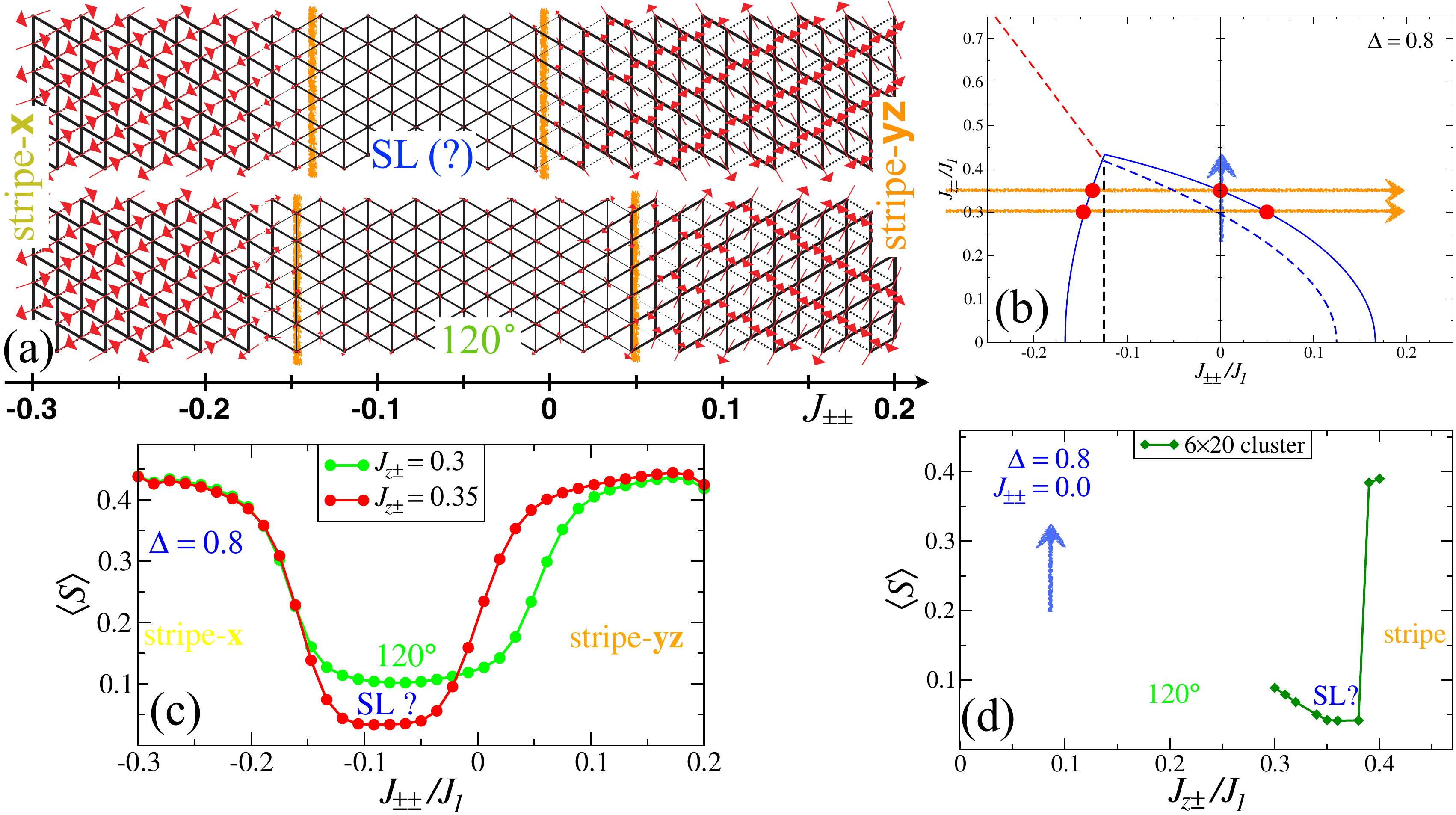}
\caption{Same as in Fig.~\ref{sFig6D1} for $\Delta\!=\!0.8$. 
(a) The scans parallel to the $J_{\pm\pm}$-axis at $J_{z\pm}\!=\!0.3$ and 0.35 as is shown in (b) 
[c.f. Fig.~\ref{sFig6D1}(b)].
(c) The  $\langle S\rangle$ vs $J_{\pm\pm}$  profiles from the scans in (a) [c.f. Fig.~\ref{sFig6D1}(e)]. 
(d) The  $\langle S\rangle$ vs $J_{z\pm}$  profile from $6\!\times \!20$ cylinders for the cut 
along the  $J_{z\pm}$-axis [c.f. Fig.~\ref{sFig6D1}(d)].}
\label{sFig7D08}
\end{figure*}

\emph{Phase diagram, $\Delta\!=\!0.8$.}---%
We continue  investigation of the model (\ref{sH}), (\ref{sHbd}) in the plane of 
$\Delta\!=\!0.8$ with scans similar to the ones in Figs.~\ref{sFig6D1}(b) and (d). 
Our Fig.~\ref{sFig7D08}(a) presents the real-space images of the  long-cylinder scans
along the cuts parallel to the $J_{\pm\pm}$-axis at $J_{z\pm}\!=\!0.3$ and 0.35, 
indicated in Fig.~\ref{sFig7D08}(b), same as in Fig.~\ref{sFig6D1}.
Fig.~\ref{sFig7D08}(c) shows $\langle S\rangle$ vs $J_{\pm\pm}$ profiles for these cuts
and  Fig.~\ref{sFig7D08}(d) shows the scan by the $6\!\times \!20$ cylinders with fixed parameters
along the  $J_{z\pm}$-axis at $J_{\pm\pm}\!=\!0$ for the region near transition to the stripe phase.

It is useful to compare with the results for $\Delta\!=\!1.0$ in Fig.~\ref{sFig6D1}.
First, the expansion of the $120{\degree}$ state beyond the classical phase boundaries is still present, 
but is significantly smaller; see Fig.~\ref{sFig7D08}(b). 
Second, the $J_{z\pm}\!=\!0.3$ cut in  Figs.~\ref{sFig7D08}(a) and (c) shows 
that the $120{\degree}$ state reclaims its territory from the spin-liquid state. Note that there are no 
boundary conditions applied at the ends of long cylinders where a stripe state is expected, so the appearance 
of  the $120{\degree}$ state in the middle of the $J_{z\pm}\!=\!0.3$ scan is entirely spontaneous.
The $J_{z\pm}\!=\!0.35$ cut shows an intermediate SL state between the two stripe phases as before,
but in a narrower region. Lastly, the fixed-parameter, smaller cylinder scan in  Fig.~\ref{sFig7D08}(d)
still shows a kink-like feature and a flat $\langle S\rangle$ in the
narrow intermediate phase from  $J_{z\pm}\!\approx\!0.35$ to 0.38, also suggesting a much
shrunk ``suspected SL'' state. 

One can  see that the suspected SL region clearly retreats as $\Delta$ is reduced from the 
isotropic limit of the bond-independent, $XXZ$ part of the model (\ref{sH}).  
The extrapolations of the boundaries of the SL region 
indicate that it disappears completely at about $\Delta\!\approx\!0.7$.

\begin{figure*}
\includegraphics[width=1\linewidth]{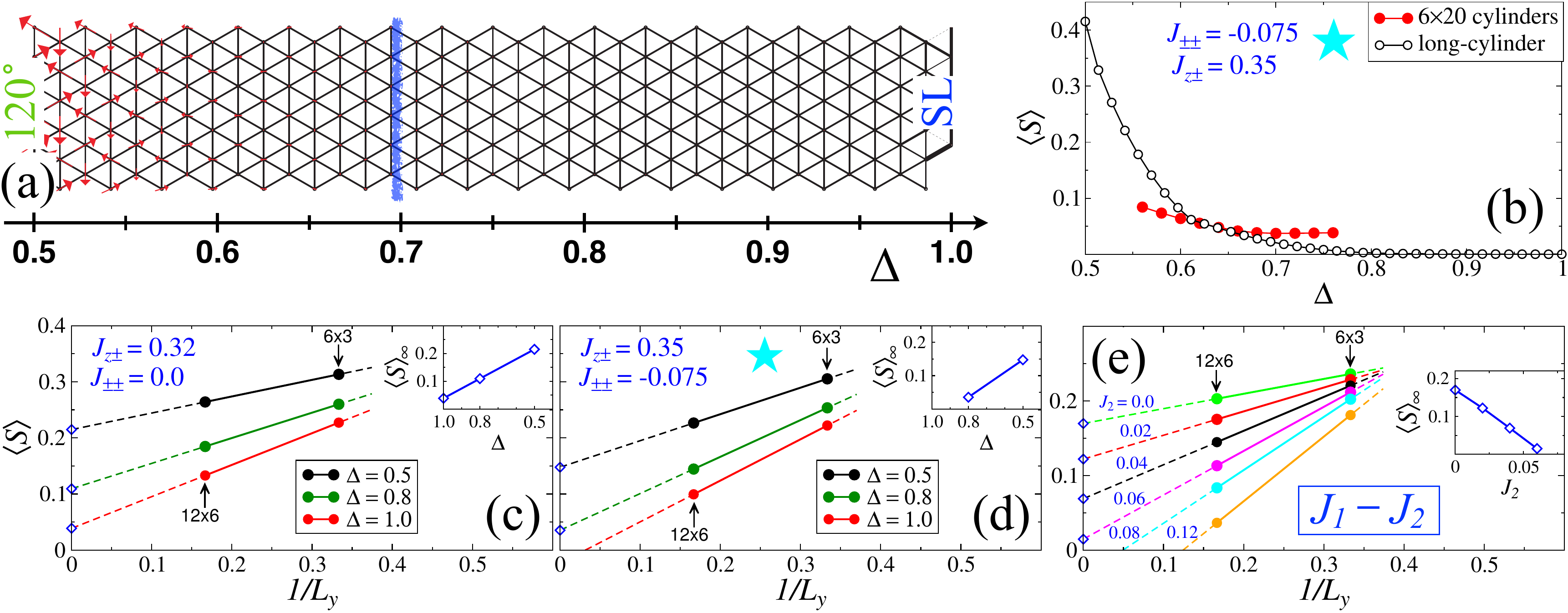}
\vskip -0.2cm
\caption{(a) and (b) 
The long-cylinder 1D DMRG scan and $\langle S\rangle$  profile at 
$[J_{\pm\pm},J_{z\pm}]\!=\![-0.075, 0.35]$ along the $\Delta$-axis, respectively. In (b), results from 
$6\!\times \!20$ cylinders are also shown.
(c) and (d) The $1/L_y$ extrapolation of $\langle S\rangle$ in the nearest-neighbor anisotropic 
triangular-lattice model (\ref{sH}), (\ref{sHbd})
for $\Delta\!=\!1.0$, 0.8, and 0.5 and for $[J_{\pm\pm},J_{z\pm}]\!=\![0.0, 0.32]$ [(c)] 
and for $[J_{\pm\pm},J_{z\pm}]\!=\![-0.075, 0.35]$ [(d)].
The insets show extrapolated values $\langle S\rangle_\infty$  vs $\Delta$.
(e) The same for the $J_1$--$J_2$ model, inset shows extrapolated values of $\langle S\rangle_\infty$  vs $J_2$.}
\label{sFigDelta}
\vskip -0.3cm
\end{figure*}

\emph{Phase diagram, scan vs $\Delta$.}---%
The last statement is verified by an additional 1D DMRG scan from the middle of the 
suspected SL region in the $\Delta\!=\!1.0$ plane  at $J_{\pm\pm}\!=\!-0.075$ and $J_{z\pm}\!=\!0.35$
(marked by a turquoise star in Fig.~2 of the main text) 
along the $\Delta$-axis toward a well-formed  $120{\degree}$ region at $\Delta\!=\!0.5$.
The results are shown in Fig.~\ref{sFigDelta}(a) and (b). The boundary 
on the SL side  is open (no boundary conditions)
with one site removed to avoid spinon localization, common to $Z_2$
spin-liquid states \cite{sZhuWhite}. 
The $120{\degree}$ order parameter profile shows a transition to a magnetically 
disordered state at $\Delta\!\agt\!0.7\pm 0.05$, confirming our estimate from the extrapolation above.
In the long-cylinder profile of $\langle S\rangle$, there is an additional kink at $\Delta\!\agt\!0.62$,
suggestive of an intermediate ordered state.

\emph{$1/L$ extrapolations.}---%
Yet another test of the SL region is provided by the $1/L$ scaling of the order parameter 
$\langle S\rangle$, measured at the center of the cluster, using a sequence of clusters 
with the fixed aspect ratio  \cite{swhite07}.
In Figs.~\ref{sFigDelta}(c) and (d), we reproduce  for clarity two of such $1/L$ extrapolations from Fig.~3 of 
the main text. They show $\langle S\rangle$ vs $1/L_y$ using clusters $3\!\times\!6$ and $6\!\times\!12$
for two sets of $J_{\pm\pm}$ and $J_{z\pm}$ from the suspected SL region for $\Delta\!=\!1.0$ and 
also for smaller $\Delta\!=\!0.8$ and 0.5. The extrapolated values of the order parameter 
$\langle S\rangle_\infty$ are shown on the 
vertical axes of Figs.~\ref{sFigDelta}(c) and (d) and plotted in their insets as a function of $\Delta$. 
This analysis indicates that the SL region of the 3D parameter space is likely much smaller than  
found by the cylinder scans. For instance, the $1/L$ extrapolation demonstrates a weak order along the
$J_{z\pm}$-axis even for $\Delta\!=\!1.0$,  see Fig.~\ref{sFigDelta}(c), and the extent of the 3D SL region
along the $\Delta$-axis is limited by $\Delta\!\approx\!0.9$, see Figs.~\ref{sFigDelta}(d), not by $\Delta\!\alt\!0.7$
as was found earlier.

This dichotomy of the results from the complementary methods must be contrasted with the 
case of the other well-studied spin-liquids, such as the isotropic $J_1$--$J_2$ model on the triangular lattice
\cite{sZhuWhite}. In Fig.~\ref{sFigDelta}(e) we demonstrate the $1/L$ scaling for this model
using the same set of clusters for several values of $J_2$  (in units of $J_1$). 
The extrapolations and the plot of  $\langle S\rangle_\infty$  vs $J_2$ in the inset show a transition
to an SL state at $J_2\!\agt\!0.06$, the value that is in an almost precise agreement with several other
criteria, such as the long-cylinder scans, correlation length decay, and energy extrapolation \cite{sZhuWhite,s_us}.
This is not the case in the present study, and while the exact reason for such a lack of a close agreement by 
different methods is unclear, we suspect that this might be another indication of a remnant or a competing order.
 
\emph{Summary of the nearest-neighbor model.}---%
With the help of the data described above, we are able to construct an approximate phase diagram
of the quantum, $S=1/2$, nearest-neighbor anisotropic triangular-lattice model in (\ref{sH}), (\ref{sHbd}),
see Fig.~4 of the main text, in which we give a generous outline to the spin-liquid state region according
to the results of the long-cylinder scans. The SL state forms a deformed cone-like shape in a proximity of the line 
where all three classical phases, the $120{\degree}$, the stripe-{\bf x}, and the stripe-{\bf yz} states meet with a 
limited extent along the $\Delta$-axis and into the classical $120{\degree}$ region.
From this analysis it is clear that the suspected SL state has the largest footprint in the $\Delta\!=\!1.0$ 
plane and is, thus, favored by a more isotropic form of the $XXZ$ part of the model (\ref{sH}), 
contrary to the expectations that the anisotropic terms in (\ref{sHbd}) are the ones that are driving the system to a 
massive degeneracy that generates an SL state. 
This feature also hints at a possible connection with the other forms of spin-liquid states in an extended phase 
diagram of the model  that we explore next.

\vspace{-0.1cm}
\subsection{$J_2$-extension}

\begin{figure*} 
\includegraphics[width=1\linewidth]{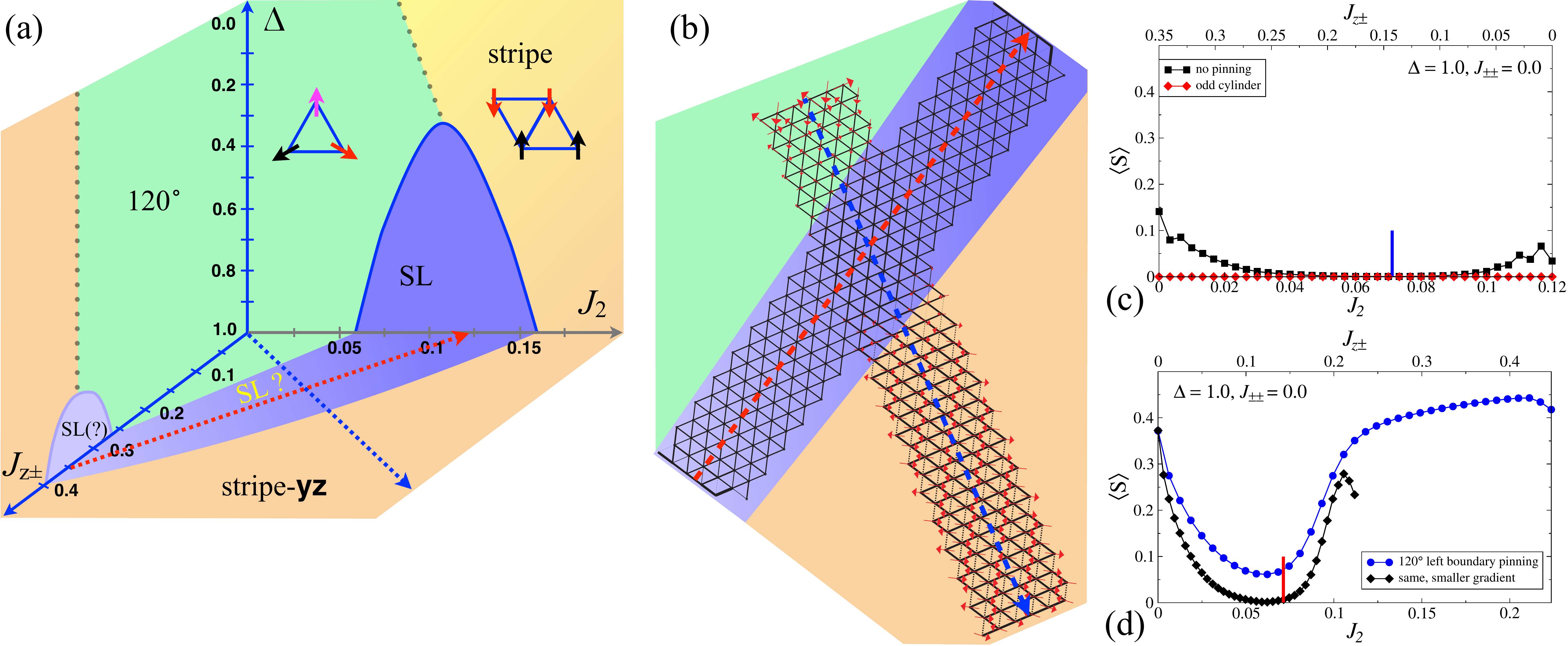}
\caption{(a) The 3D, ($J_{z\pm}, J_2,\Delta$) phase diagram of the extended 
anisotropic model with $J_2$ at $J_{\pm\pm}\!=\!0$.
The l.h.s. and r.h.s. panels are the 2D $J_{\pm\pm}\!=\!0$  section of the 3D diagram 
of the nearest-neighbor model and the 2D phase diagram of the $XXZ$ $J_1$--$J_2$ model, respectively.
Arrows show directions of the 1D long-cylinder DMRG cuts in (b). (b) The real-space images of cylinders 
with arrows showing directions of the scans and phases marked schematically.
(c) and (d) The  $\langle S\rangle$ vs $J_2 (J_{z\pm})$  profiles from (b) with different boundary conditions
or parameter range, see text. Vertical bars mark the intersect of the scans.}
\label{sFig9}
\vskip -0.3cm
\end{figure*}

There are two reasons to be interested in a wider parameter space 
that extends the nearest-neighbor anisotropic triangular-lattice model of (\ref{sH}), (\ref{sHbd}).
First,  the experiments in YbMgGaO$_4$  (YMGO), one of the most-studied experimental 
realization of such a model,  strongly suggest a sizable 
second-nearest-neighbor $J_2$ term \cite{sMM}. Second, the spin liquid in 
the isotropic $J_1$--$J_2$ model on the triangular lattice is well studied \cite{sZhuWhite},
so it would be very informative to establish a relation between the spin liquids in these models, if it
exists.

For both reasons, a minimalistic modification of the nearest-neighbor model (\ref{sH}), (\ref{sHbd})
by the $XXZ$-only next-nearest-neighbor $J_2$-term suffices
\begin{eqnarray}
\label{sHJ2}
{\cal H}_{XXZ}^{J_2}=J_2\sum_{\langle\langle ij\rangle\rangle} 
\left(S^{x}_i S^{x}_j+S^{y}_i S^{y}_j+\Delta S^{z}_i S^{z}_j\right),
\end{eqnarray}
where we also assume that the $XXZ$ anisotropy is the same as in the $J_1$-term in (\ref{sH}).

There are several ways of representing the four-dimensional parameter 
space of the resultant extended model.  
One is offered in our Fig.~4 of the main text, in which the back panel represents a topographic 
map of the nearest-neighbor model and the bottom panel is such a map 
for the $XXZ$ $J_1$--$J_2$--$J_{\pm\pm}$ (or anisotropic $J_1$--$J_2$, $J_{z\pm}\!=\!0$) model,
where the topographic axis for both panels is the $XXZ$ anisotropy $\Delta$.
For the latter model, we used the results obtained by us previously in Ref.~\cite{s_us}.  
Such a ``dual'' map is suggestive of the connectivity between the corresponding 2D ``puddles'' of the 
SL states  in the two panels  via some $J_{z\pm}$--$J_2$ ``tube'' for a fixed  $\Delta$.

\begin{figure*}[t]
\includegraphics[width=1\linewidth]{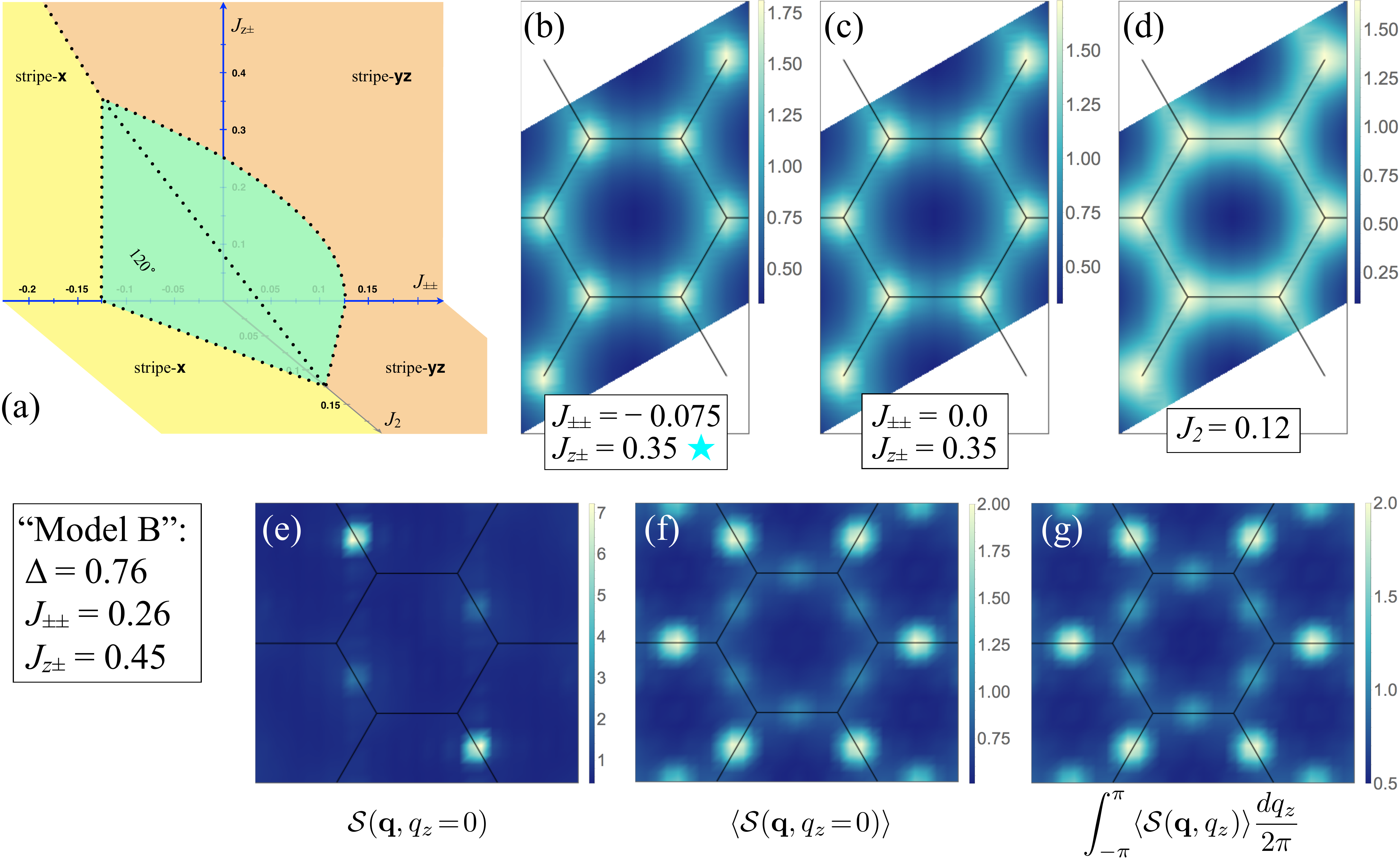}
\caption{(a) The classical 3D phase diagram of the $J_2$-extended model at $\Delta\!=\!1.0$.
(b)-(d) Static structure factor, ${\cal S}({\bf q})$,  from the $6\!\times \!20$ cylinders
for the anisotropic nearest-neighbor model at $\Delta\!=\!1.0$ at 
$(J_{\pm\pm},J_{z\pm})\!=\!(-0.075, 0.35)$ [(b)] and 
$(J_{\pm\pm},J_{z\pm})\!=\!(0.0, 0.35)$ [(c)], and for the isotropic $J_1$--$J_2$ model at $J_2\!=\!0.12$ [(d)].
(e)-(g) ${\cal S}({\bf q})$ for the stripe state, ``Model B'' parameters, at $q_z\!=\!0$ [(e)],
averaged over three domain orientations [(f)], and integrated over $q_z$ [(g)].}
\label{sFig10}
\vskip -0.3cm
\end{figure*}

It is also natural to expect such a connection from the purely classical 
phase diagram. For a fixed $\Delta$, the 3D parameter space is given by the ($J_2,J_{\pm\pm}, J_{z\pm}$) triad
and the region occupied by the $120{\degree}$ state is
a deformed tetrahedron, with an example of the phase diagram for $\Delta\!=\!1.0$
shown in Fig.~\ref{sFig10}(a). The back and the bottom panels of it are the same as in Fig.~4 of the main text.
Since the SL states in both 2D panels occur in a proximity of a $120{\degree}$-to-stripe-to-stripe tricritical point, 
then the connection, if it exists, should be expected along the $J_{z\pm}$--$J_2$ tricritical rim of the 
$120{\degree}$ state tetrahedron in Fig.~\ref{sFig10}(a). 

A somewhat different 3D ``cut" of the 4D parameter space is offered in Fig.~\ref{sFig9}(a).
Instead of fixing $\Delta$, we select a plane of $J_{\pm\pm}\!=\!0$ in Fig.~\ref{sFig10}(a) and add
$\Delta$ as the third axis, with the resultant triad  being ($J_{z\pm}, J_2,\Delta$). Such a choice is motivated by the 
fact that $J_{\pm\pm}$ is detrimental to the $J_1$--$J_2$ spin-liquid state according to our previous
study \cite{s_us}.
Thus, the left-hand-side panel in Fig.~\ref{sFig9}(a) is the $J_{\pm\pm}\!=\!0$ section of the 3D diagram 
of the nearest-neighbor model and the right-hand-side one is the phase diagram of the $XXZ$ $J_1$--$J_2$ model.

The isotropic $J_1$--$J_2$ model ($\Delta\!=\!1.0$) has a spin-liquid ground state for a range of 
$J_2$ between 0.06 and 0.16 (units of $J_1$) \cite{sZhuWhite}. As we have shown in our previous study \cite{s_us},  
it survives the $XXZ$ anisotropy down to $\Delta\!\approx\!0.3$; see Fig.~\ref{sFig9}(a).
One can see that both l.h.s. and r.h.s. panels in Fig.~\ref{sFig9}(a) have domes of an SL state. 
A natural test of their compatibility is provided using our 1D DMRG  long-cylinder scans shown 
schematically by arrows in Fig.~\ref{sFig9}(a), see also Fig.~4 of the main text.
For the ($J_{z\pm}, J_2,\Delta$) coordinates, the scans are along the     
$(0.35, 0.0, 1.0)\!\rightarrow\!(0.0, 0.12,1.0)$ and $(0.0, 0.0, 1.0)\!\rightarrow\!(0.44, 0.22,1.0)$
directions. The results are presented in Figs.~\ref{sFig9}(b), (c) and (d), with the vertical bars in 
Figs.~\ref{sFig9}(c) and (d) showing the intersection point of the scans, 
see also Fig.~5(a) and (b) of the main text.

The first of the scans connects anisotropic spin-liquid state  
identified in Fig.~\ref{sFig6D1} with the isotropic $J_1$--$J_2$ spin liquid known
previously \cite{sZhuWhite}, from  $(J_{\pm\pm},J_{z\pm})\!=\!(0.0, 0.35)$ to $J_2\!=\!0.12$.
We have tried two types of boundary conditions as Fig.~\ref{sFig9}(c) indicates. First is just
open boundaries. Second is the same, but with one site removed at each end to suppress 
spinon localization at the edges \cite{sZhuWhite}. While the first cylinder develops a weak order at the edges, the
second scan shows no indication of a magnetic order anywhere along its length, 
clearly demonstrating that the SLs of the two models are connected. More importantly, this scan displays 
the same  thickness of the bonds along the cylinder, which is proportional to the nearest-neighbor correlation 
$\langle {\bf S}_i{\bf S}_j\rangle$, thus showing no change of the character of the SL state anywhere along this scan.

The second scan starts at the Heisenberg, nearest-neighbor point ($J_1$-only model) and extends deep into the 
stripe-{\bf yz} state. It is complementary to the first scan and is also used to confirm 
the existence of an intermediate SL state between the $120{\degree}$ and the stripe phases 
along the direction tilted from both $J_2$ and $J_{z\pm}$ axes.  
Fig.~\ref{sFig9}(d) shows two profiles of $\langle S\rangle$ vs $J_2 (J_{z\pm})$, one for the entire range of parameters
and one for a shorter range and smaller gradient. It clearly demonstrates the robust 
intermediate SL state, consistent with the first scan as well as with the scans of the same nature along 
$J_{z\pm}$ and $J_2$ axes in Fig.~\ref{sFig6D1} and in Ref.~\cite{s_us}.

\vspace{-0.3cm}
\subsection{Static structure factor ${\cal S}({\bf q})$}

While the nearest-neighbor correlator $\langle {\bf S}_i{\bf S}_j\rangle$ 
has already demonstrated no change of the character of the 
SL state anywhere along the long-cylinder scan in Fig.~\ref{sFig9}, 
a deeper and more comprehensive insight into the nature of the SL state is offered by 
the static structure factor, ${\cal S}({\bf q})\!=\!\sum_{\alpha\beta}(\delta_{\alpha\beta}-q_\alpha q_\beta/q^2)
{\cal S}^{\alpha\beta}_{\bf q}$, a quantity that is also observable in the neutron-scattering experiments.
We obtain it from the Fourier transform,
${\cal S}^{\alpha\beta}_{\bf q}\!=\!N^{-1}\sum_{ij}\langle S_i^\alpha S_j^\beta\rangle e^{i{\bf q}({\bf R}_i-{\bf R}_j)}$,
of the real-space spin-spin correlation function $\langle S_i^\alpha S_j^\beta\rangle$,
where the latter is determined from the ground state wave-function on various DMRG cylinders.

In the main text, we have 
used  correlation function from the different section of the long cylinder in Figs.~\ref{sFig9}(b),(c), see also 
Fig.~5  of the main text  and its discussion. 
Here we show ${\cal S}({\bf q})$ from the $6\!\times \!20$ cylinders with fixed parameters
for three representative sets of data; see Figs.~\ref{sFig10}(b)-(d). 
In Figs.~\ref{sFig10}(b) and (c),  ${\cal S}({\bf q})$ is for the
anisotropic nearest-neighbor model  at $\Delta\!=\!1.0$ and for  
$(J_{\pm\pm},J_{z\pm})\!=\!(-0.075, 0.35)$ and $(J_{\pm\pm},J_{z\pm})\!=\!(0.0, 0.35)$, respectively.
 In Fig.~\ref{sFig10}(d),  ${\cal S}({\bf q})$ is for the isotropic $J_1$--$J_2$ model at $J_2\!=\!0.12$.
Thus,  ${\cal S}({\bf q})$ in Figs.~\ref{sFig10}(c) and (d) correspond to the beginning and the end 
of the long-cylinder scan in Fig.~\ref{sFig9} marked by the red arrow. 
The ${\cal S}({\bf q})$ in Fig.~\ref{sFig10}(b) is also from the SL region of  
the nearest-neighbor anisotropic triangular-lattice model of (\ref{sH}), (\ref{sHbd}); see Fig.~\ref{sFig6D1},
Figs.~2 and 3 of the main text,  and
Fig.~\ref{sFig7D08}, where it is marked by a star symbol.
 
\begin{figure*}[t]
\includegraphics[width=1\linewidth]{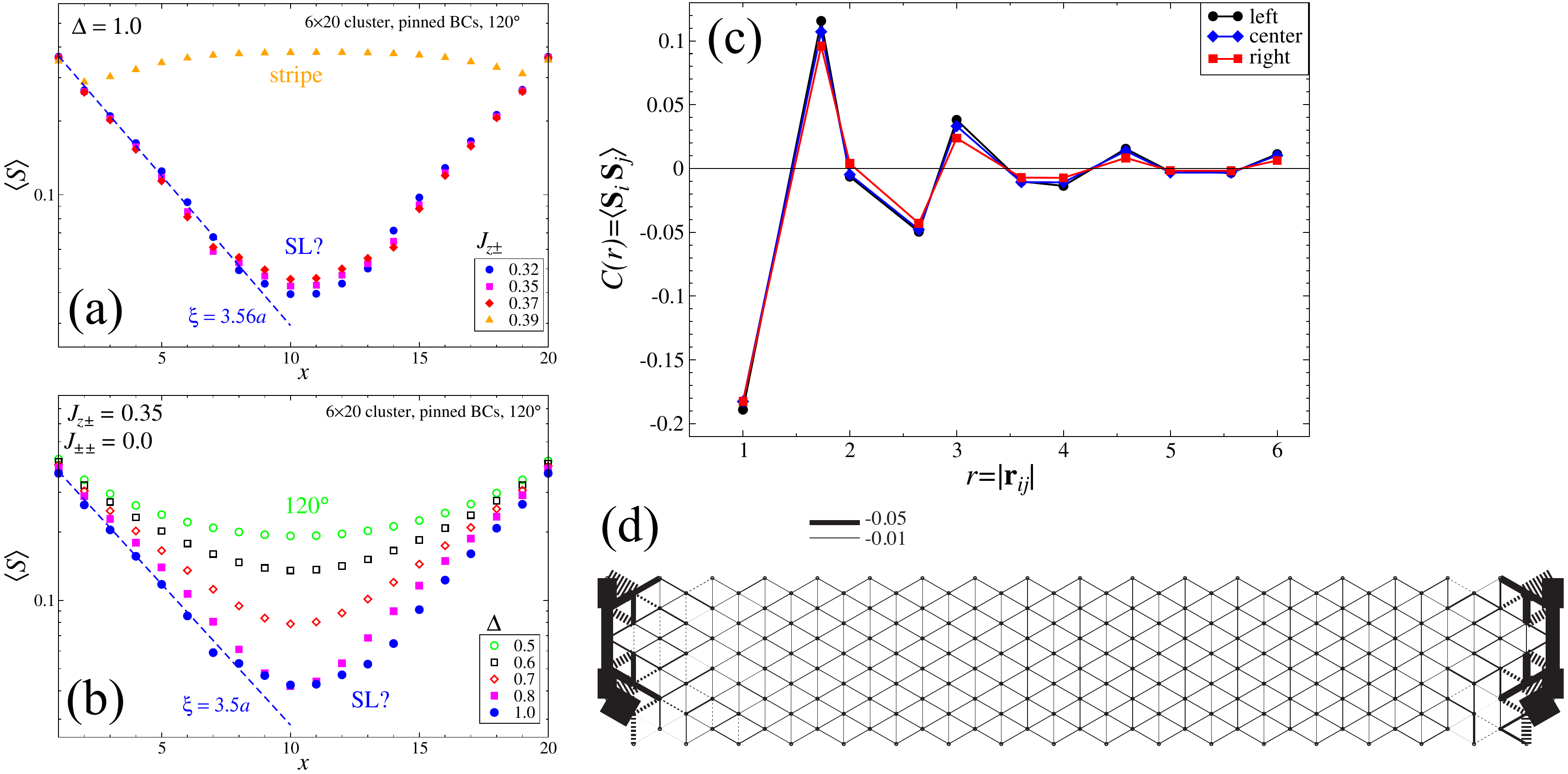}
\caption{(a) and (b) $\langle S\rangle$ vs the length of the cylinder 
in the $6\times 20$ clusters with the 120$^{\degree}$ boundary conditions
on a semi-log plot. The exponential decay in the spin-liquid state with the correlation length $\xi$ is emphasized.
(a) is for  $\Delta=1.0$, $J_{\pm\pm}=0$ for several $J_{z\pm}$. (b) is for  
$J_{\pm\pm}=0$, $J_{z\pm}=0.35$ for several values of $\Delta$.
(c) The real space spin-spin correlation function, $C(r)=\langle {\bf S}_i{\bf S}_j\rangle$, with $r=|{\bf r}_{ij}|$
in the three sections of Fig.~5(a) of the main text.
(d) The 1D scan from Fig.~5(a) of the main text showing magnified
bonds of $\langle {\bf S}_i{\bf S}_j\rangle+0.175$.}
\label{sFig11}
\vskip -0.3cm
\end{figure*}
 
 One can see that while there is some minor variation in intensity and overall scale,
 the static structure factor is essentially the same for all three
 parameter sets, confirming the results inferred from the long-cylinder scan.
This implies, once more, that the SLs  in the anisotropic nearest-neighbor model and   
in the fully isotropic $J_1$--$J_2$ model, as well as in any interpolating SL state, are isomorphic to each other.

Of the discussed models, the  isotropic $J_1$--$J_2$ model is most studied \cite{sZhuWhite}, see the main
text for a more comprehensive list of references. The exact nature of its spin-liquid state is still under  debate,
with the broadened peak in ${\cal S}({\bf q})$ at the $K$-points consistent with the nematic
$Z_2$ \cite{sZhuWhite}, $U(1)$ Dirac,  chiral, and Dirac-like spin-liquids. 
One thing is clear: it is \emph{not} consistent with the ``open spinon Fermi surface'' SL state 
proposed for YMGO \cite{sChen2}, see also Ref.~\cite{sBalents17}. 
This observation alone is sufficient to suggests that an extrinsic mechanism is responsible 
for an SL-like response in this material.

Three additional points are in order. In the spin-liquid states in Figs.~\ref{sFig10}(b)-(d), 
${\cal S}({\bf q})$ is essentially isotropic, meaning that the contribution of the off-diagonal terms in it,
${\cal S}^{\alpha\neq\beta}_{\bf q}$, are negligible in our results, a feature naturally 
expected for a magnetically isotropic,  spin-liquid-like state. In addition, there is  only a very weak
$q_z$ dependence of ${\cal S}({\bf q})$ and the broadened peaks are 
obviously the same in different Brillouin zones.

Consider now a stripe state. For clarity, we choose the ``Model B'' set of parameters 
discussed in Figs.~\ref{sFig4D05}(b) and (d). Our Fig.~\ref{sFig10}(e) shows ${\cal S}({\bf q})$ for it
at $q_z\!=\!0$ and Fig.~\ref{sFig10}(f) shows the same data averaged over three possible stripe orientations.
The last Figure \ref{sFig10}(g) is the same integrated over $q_z$.
First, experiments in YMGO \cite{sMM,sChen2} show the structure factor with the maxima of intensity at the 
$M$-points, a feature easily obtainable from a mixture of stripe domains, see Fig.~\ref{sFig10}(f),
supporting a scenario of the disorder-induced spin-liquid mimicry in YMGO, proposed by us in Ref.~\cite{s_us}.

In addition, we have verified that: (i) the role of the 
off-diagonal terms in ${\cal S}({\bf q})$ of the stripe phase
in  Figs.~\ref{sFig10}(e)-(f) is substantial, (ii) ${\cal S}({\bf q})$ has a significant $q_z$ variation, and (iii)
there is a clear and substantial variation of the peaks at the $M$-points from one Brillouin zone to the other.
We note that the latter feature has been clearly observed in the neutron-scattering experiments in 
YMGO \cite{sMM}, yielding a yet more support to the disordered stripe state scenario \cite{s_us}.

Lastly, we suggest that the polarized neutron scattering may be able to 
identify contribution of the off-diagonal ${\cal S}({\bf q})$ components in YMGO.

\vspace{-0.3cm}
\subsection{Correlations and other orders}

\emph{Correlation length.}---%
For the suspected spin-liquid region of the model (\ref{sH}) and (\ref{sHbd}),
we have measured the decay of the ordered moment $\langle S\rangle$ away from the 
boundaries  in the $6\times 20$ clusters with the 120$^{\degree}$ boundary conditions. In the 
spin-liquid state, such decays are always exponential with the correlation length
about $3-4$ lattice spacings, see Fig.~\ref{sFig11}(a) and (b). The first plot shows the
ordered moment $\langle S\rangle$ vs the length of the cylinder on a semi-log plot to emphasize 
the exponential form of  the decay. It is for fixed $\Delta=1.0$ and $J_{\pm\pm}=0$
and for several values of  $J_{z\pm}$ from the spin-liquid region with one set from the neighboring stripe phase
for a contrast. The second panel shows the same type of data for  
$J_{\pm\pm}=0$ and  $J_{z\pm}=0.35$ and for several values of $\Delta$, also confirming the results in 
Fig.~\ref{sFigDelta} that a transition from the 120$^{\degree}$ order to an SL state is near 
$\Delta\sim 0.7-0.8$.

This is yet another type of  measurement that we use to confirm the existence of the spin-liquid phase.
One of the problems with using it to identify the extent of a spin-liquid phase  
is that the decay  is still exponential when the order is weak but is certainly present, thus leading 
to an overestimate of such a region.

\emph{Correlation function of the spin-liquid state.}---%
In addition to the structure factor $S({\bf q})$ above, we provide further support to the isomorphism 
of the spin-liquid phase throughout $J_1$--$J_2$--$J_{z\pm}$ parameter space,
by investigating the real space spin-spin correlation 
function, $C(r)=\langle {\bf S}_i{\bf S}_j\rangle$, where $r=|{\bf r}_{ij}|$.
We have chosen the same three regions of the spin-liquid phase in the ``spin-liquid cylinder'' of 
Fig.~5 of the main text [also shown in Fig.~\ref{sFig9}], for which we had the nearly identical $S({\bf q})$ maps.

The results for $C(r)$ for   three different sections of the cylinder, marked
 ``left,'' ``center,''  and ``right,'' are shown in Fig.~\ref{sFig11}(c). The correlations are measured, roughly,
along the diagonal of each of the  section and reach the distances $r\!=\!6a$ that corresponds to  the 11th
nearest-neighbor. One can plainly see that even the fine structure of the correlations is the same, e.g., showing 
a near-zero correlation for the third neighbor, and that $C(r)$'s remain 
quantitatively close all the way from the anisotropic spin-liquid to the isotropic  $J_1$--$J_2$ one. 
Therefore, according to our measurements, with the accuracy one can detect, the 
``anisotropic'' spin liquid and ``isotropic'' spin liquids are the same, or isomorphic to each other.

\emph{Valence-bond and chiral orders.}---%
Although we have not observed any indication
of the dimerized states in any of our scans [the thickness of the bonds 
in all our cylinders is proportional to the nearest-neighbor correlation 
$\langle {\bf S}_i{\bf S}_j\rangle$], we have also conducted a direct search
for possible dimerization patterns, see Fig.~\ref{sFig11}(d).  
This Figure shows the same 1D ``spin-liquid cylinder'' scan as in Fig.~5(a) of the main text  and above
with the value $-0.175$ subtracted from the nearest-neighbor correlation $\langle {\bf S}_i{\bf S}_j\rangle$ 
(the average value of it is about $-0.18$). The result of the subtraction is magnified to make the subtler 
variations visible on the scale of 0.01 or less. Aside from the open-boundary-related effects, 
no sign of a dimerization pattern is detected.

For the same 1D scan, we have searched for a non-zero chirality by measuring 
$\langle {\bf S}_1\left({\bf S}_2\times{\bf S}_3\right)\rangle$ on each triangle and found that 
the chirality values are less than 10$^{-8}$, i.e., beyond detectability.

\vspace{-0.3cm}



\begin{thebibliography}{99}
\bibitem{footnote1} For a less humane form of (\ref{H}), see, e.g., \cite{Chen1}.
We use a standard form of the $XXZ$ term (cf.  \cite{Chen1} and others) 
with $J\!=\!2{\sf J_\pm}$ and anisotropy $\Delta\!=\!{\sf J_z}/2{\sf J_\pm}$ in their notations and 
a different operator form in ${\cal H}_{\rm bd}$ (\ref{H}).
\bibitem{ModelB}  For $\Delta\!=\!0.76$, $J_{\pm\pm}\!=\!0.26J$, $J_{z\pm}\!=\!0.45J$ (${\sf J_\pm}\!=\!0.66{\sf J_z}$, 
${\sf J_{\pm\pm}}\!=\!0.34{\sf J_z}$, ${\sf J_{z\pm}}\!=\!0.6{\sf J_z}$ in the notations of Ref.~\cite{Chen4}),
ordered moment is $\langle S\rangle\!\approx \! 0.4694 (0.4763)$
by DMRG  (SWT).
\bibitem{footnoteSq} No significant variation is detected vs $q_z$.
\end{thebibliography}
\end{document}